\documentclass[aps,a4paper,twocolumn,floatfix,showkeys,
groupaddress]{revtex4}

\usepackage{bm,amsmath,amssymb,amsfonts}

\usepackage[dvips]{graphicx}
\usepackage[colorlinks=true,citecolor=magenta,urlcolor=blue,filecolor=red]{hyperref}
\usepackage{color}
\definecolor{mycolor}{rgb}{0.7,0.3,0}
\begin{document}

\title{\textcolor{mycolor}
{Unified strategy of flat band engineering, electronic transport and other spectral properties
for different kagom\'{e} ribbon variants}}  

\author{Atanu Nandy}
\email{atanunandy1989@gmail.com}
\affiliation{Department of Physics, Acharya Prafulla Chandra College, New Barrackpore, Kolkata,
West Bengal-700 131, India}

\begin{abstract}
We address the problem of flat band engineering in different prototypes of quasi-one dimensional kagom\'{e} network through a generalized analytical proposition worked out within the tight-binding formalism. 
Exact fabrication of single particle eigenstates with localized as well as diffusive modes is reported through the demonstration of such unified methodology by virtue of
 a simple real space decimation formalism in such interesting variants of ribbon shaped geometry.
The description provides a common platform to investigate
the band dispersion including the overall spectral portrait and associated physical aspects
 of those quasi-one dimensional lattices. 
 Exact detection of dispersionless flat band mode and its tunability are reported as a direct consequence of the analytical
 prescription.
%This scheme enables us to study the flat band engineering for all the members considered. 
 Analytical work out is justified through the numerical evaluation of density of eigenstates, electronic transmission behavior, inverse participation ratio, persistent
 current study, Aharanov-Bohm oscillation in the transmittance and other related issues.
 An obvious analogous extension in the context photonics concludes our description.
\end{abstract}
%\pacs{71.30.+h, 72.15.Rn, 03.75.-b}
\keywords{Flat band, Aharonov-Bohm oscillation, persistent current, quantum butterfly.}
\maketitle
%%%%%%%%%%%%%%%%%%%%%%%%%%%%%%%%%%%%%%%%%%%%%%%%%%%%%%%%%%%%%%%%%%%%%%%%%%%%%%%%%%
\section{Introduction}
\label{intro}
A momentum independent flat band~\cite{suther1}-\cite{flach5} is meticulously non-dispersive in the entire Brillouin zone. The incoming electron having flat band energy behaves as heavily massive one and thus loses its movability due to the destructive kind of wave interference resulted from the network. The divergent density of states as a consequence of vanishing band curvature leads to anomalous transport behavior.
It has been considered as an appropriate platform to explore the strong correlation physics owing to the complete quenching of the kinetic information.
The fascination in this flat band physics originates from the basic fact that 
a flat band is generally associated with indigenous macroscopic degeneracy and absence of transmittance.
If the degeneracy is broken by the application of small perturbation, it results
in novel transporting phases whose 
characteristic features depend on the perturbations incorporated. 
Thus, the perturbations play a pivotal role, and that is 
what turns flat bands into momentous starting points for 
switching between different phases of matter.
It is well-known that B. Sutherland first investigated the compact
localized eigenstates in tight-binding
models of quasicrystals~\cite{koho} as well as in systems with
strict discrete translational invariance~\cite{suther1}.
The subsequent observation has been exhibited in case of Lieb lattice~\cite{lieb1}. The discovery of Mielke~\cite{mie}
and Tasaki~\cite{tasaki} have also added substantial momentum in this realm.

It is needless to say that electronic spectra of several deceptively simply looking tight-binding lattices with nearest neighbor connectivity have displayed exotic characters such as the appearance of dispersionless, flat 
bands (FB)~\cite{mati,lopes}. All these frustrated hopping models are found as interesting.
Because of the exponentially large degeneracy, they offer prospects of strong interaction physics 
such as the fractional quantum Hall effect, unconventional 
inverse Anderson transition~\cite{shukla,goda}, multifractality at weak 
disorder~\cite{aldea}, Hall ferromagnetism~\cite{kimura,ueda}, etc.
Graphene anti-dot lattices~\cite{van} and also
post-graphene members like
phosphorene~\cite{eza} have been identified as networks exhibiting
the qualities for quantum computation~\cite{flin} and an electrical 
tunability of quasi-flat bands respectively.
This field of research also 
finds rising interest in search of non-trivial flat bands as they allow us to study the lattice version of fractional 
topological phenomena~\cite{neu,sun}.
Recent advancement in the lithography techniques flat bands are realized 
experimentally~\cite{seba1}-\cite{rodrigo} in the area of photonics using the standard ultrafast laser writing
 technique. All these have unfolded a valid point of discussion related to flat band systems. Several periodic~\cite{mati}, quasi-periodic~\cite{an3} and fractal models~\cite{bp} have been examined in this regard.

Kagom\'{e} lattice, the most intriguing one, has a long expedition.
It is well-known that
a spin kagom\'{e} structure is a very promising
 candidate for quantum spin liquid due to the inherent geometrical frustration associated with it.
Rigorous investigation has been done in search of spin liquid behavior in different
hybrid  kagom\'{e}  metals~\cite{iq}-\cite{chu} which are essentially the compounds containing layers of 
 kagom\'{e} sublattice of transition metals sandwiched between a couple of layers of organic ligands. 
The same network has also been well studied to find the
other magnetic quantum states such as quantum optical spin ice~\cite{barun,qi}, 
kagom\'{e} magnet~\cite{chu}, anomalous Hall effect~\cite{dee} and skyrmion~\cite{hou}.
Thus the kagom\'{e} structure has opened up different challenging domains in diverse topics of
condensed matter physics.
In recent times, inorganic kagom\'{e} metals have significantly highlighted because of the existence of Dirac bands with van Hove singularities and 
the eventual flat band, which finally 
motivates us to observe several novel physical phenomena, such as 
ferromagnetism~\cite{mie,tasa}, superconductivity~\cite{miya}.

In this article, our point of interest is the
the quasi-one dimensional kagom\'{e} ribbon~\cite{moli} 
which is instinctively
dimensionally reduced two-dimensional
 kagom\'{e} network. 
Fabrication of such ribbon shaped structure 
using femto-second laser inscription technique~\cite{miu,nol} is a very standard way of realization which
inspires us to take all the kagom\'{e} variants as our theme of interest.
Here, we describe a general analytical prescription to discern the existence of non-dispersive flat band modes and its tunability 
with an inclusive details of the entire spectral landscape and different associated issues in a single frame for a class of quasi-one dimensional kagom\'{e} ribbon variants. This will help us to investigate all the known flat band prototypes following a definite pathway.
This route map maintains its uniformity for all the variants discussed. This can guide us to unravel the hierarchical distribution of flat band modes when the unit cell carries a `decoration' by a non-translationally invariant \textit{fractal} object. We have discussed that one may have a control over the spatial extent of the \textit{quantum prison} resulting in the staggering effect of localization.
Flux dependent flat band, a special subset of Ahoronov-Bohm caging~\cite{vidal} and the oscillatory nature of the transmittance corroborate our findings. 
As a second motivation, it has been illustrated that a
deterministic but aperiodic fashion of the external perturbation can lead to produce quantum butterfly in the energy landscape of square kagom\'{e} variant. 
It is to be noted that an identical slowly varying modulation was first utilized in fabricating an aperiodic potential landscape~\cite{das} in a one-dimensional tight-binding network. The interplay
of different parameters in this modulation invites a non-trivial staggered flux variation.
In conclusion, the direct correspondence with the optical case which inspires us to discuss the relevant photonic localization is also studied along with the evaluation of photonic dispersion relation.

In what follows we demonstrate our findings. 
Section~\ref{model} tells about the
 overall description of the system with the help of tight-binding Hamiltonian.
After that Section~\ref{star} shows the evaluation of spectral property offered
by the star ribbon network and its decorated version. The role of uniform perturbation has been studied
in Section~\ref{ben}. 
Spectral response is investigated in details for square kagom\'{e} geometry
in Section~\ref{sq}. Section~\ref{photo} discusses the analogous extension to photonics.
Finally, in Section~\ref{closing}, we summarize all the results.

\section{Model system and Hamiltonian}
\label{model}
%%%%%%%%%%%%%%%%%%%%%%%%%%%%%%%%%%%%%%%%%%%%%%%%%%%%%%%%%%%
\begin{figure}[ht]
\centering
(a) \includegraphics[clip,width=7.5 cm,angle=0]{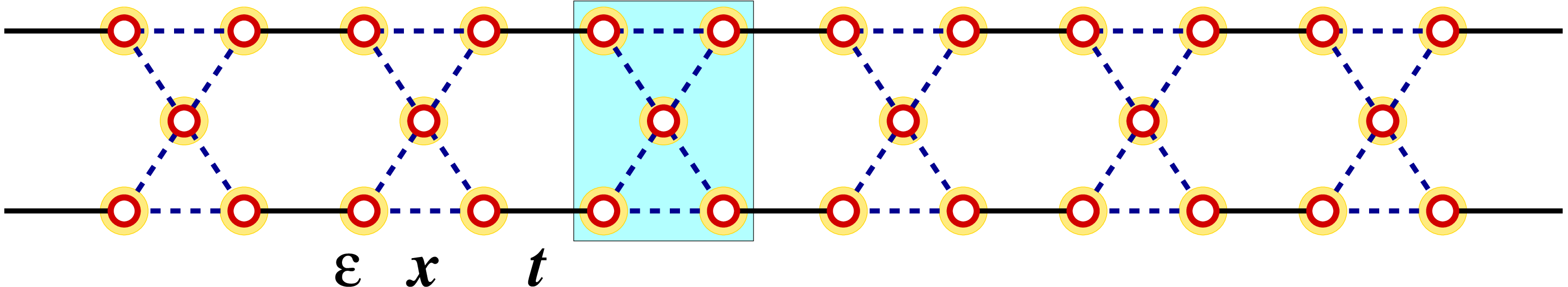}\\
(b) \includegraphics[clip,width=7.5 cm,angle=0]{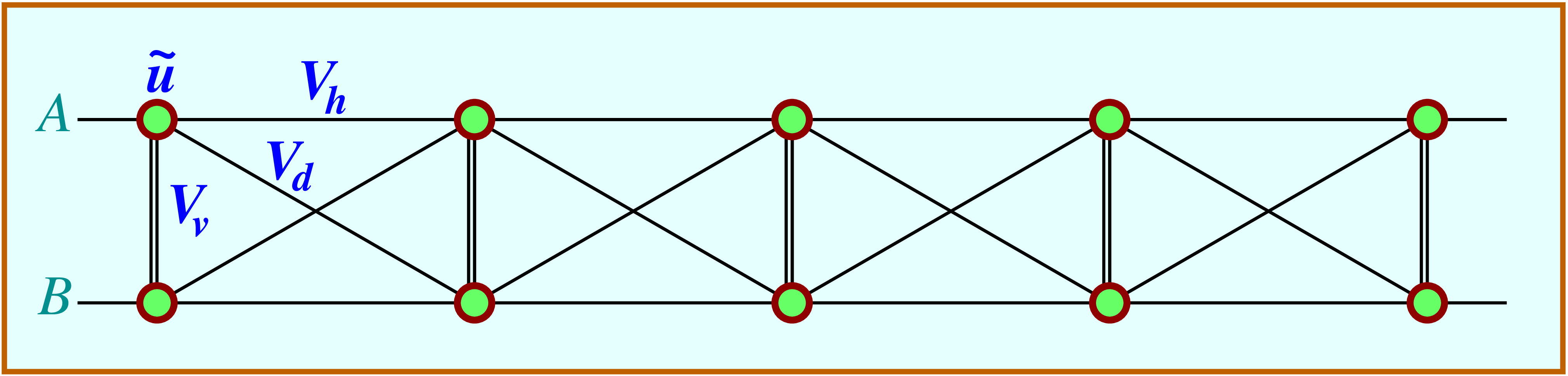}
%(c) \includegraphics[clip,width=5.3 cm,angle=0]{dens5.eps}\\
%(d) \includegraphics[clip,width=5.3 cm,angle=0]{kagome2trans1.eps}
%(e) \includegraphics[clip,width=5.3 cm,angle=0]{kagome2trans2.eps}
%(f) \includegraphics[clip,width=5.3 cm,angle=0]{kagome2trans3.eps}
\caption{(Color online) (a) A portion of an infinite kagom\'{e} ribbon network. All
the sites have the identical on-site energy $\epsilon$. The shaded region denotes the
unit cell of the structure. $x$ denotes the intra-unit cell hopping and $t$ represents the overlap
integral between two consecutive unit cells. (b) The 
\textit{effective} decimated two-arm ladder with the tight-binding parameters.}  
\label{lattice1}
\end{figure}
%%%%%%%%%%%%%%%%%%%%%%%%%%%%%%%%%%%%%%%%%%%%%%%%%%%%%%%%%%%
We discuss the 
general analytical scheme for basic kagom\'{e} ribbon (Fig.~\ref{lattice1}(a)) and
the parallel description will be extended for the other variants subsequently.
All such prototypes can be, in general, described by standard tight-binding Hamiltonian, written in the Wannier basis, viz.,
\begin{equation}
H = \sum_{\langle ij \rangle} (V_{ij} a_{i}^{\dagger} a_{j} + h.c.) +\sum_{i} u_{i} a_{i}^{\dagger} a_{i}
\end{equation}
The first part $V_{ij}$ (\textit{off-diagonal} term) denotes the overlap integral which 
contains the kinetic information of the Hamiltonian between the two consecutive atomic sites.
The later part $u_i$
(diagonal term) carries the potential energy at the respective quantum dot location.

With the basic demonstration of Hamiltonian, we will discuss the analytical pathway to discern the flat band engineering and the overall spectral landscape. At this point we should emphatically point out that this proposition, in general, works quite well for any kind of quasi-one dimensional ribbon-shaped tight-binding structure. Moreover, this analytical workout may help us to find out the information regarding the spectral property as well as the localization of single particle eigenstates offered by the underlying network in an \textit{exact} way.
We will start our analysis from the \textit{difference equation}. 
The equation which is the alternate form of Schr\"{o}dinger's equation reads as,
\begin{equation}
(E-u_{n}) \psi_{n} = \sum_{m} V_{nm} \psi_{m}
\label{diff}
\end{equation}

\subsection{Analytical scheme for kagom\'{e} ribbon}
%We will elaborately discuss the scheme here for general kagom\'{e} ribbon structure. 
The strategy
is simple and straightforward.
With the assistance of real space decimation formalism, one can simply \textit{eliminate} the wave function amplitudes of a relevant subsection of atomic sites in terms of that of the
`surviving' sites
such that the entire ribbon-like structure \textit{effectively} turns out to be a two-arm ladder (Fig.~\ref{lattice1}(b)) with energy dependent tight-binding parameters. It is to be noted that this analytical transformation is completely a mathematical mapping and the structure retains its originality intact. Hence there is no chance of loosing any physical information related to the system with the implementation of this decimation scheme. In stead, one can be acquainted with the idea of eigenspectrum using the effective parameters.
After the transformation, all the atomic sites of the two-legged ladder have now identical on-site potential $\tilde{u}$ which is essentially function of energy and the initial tight-binding parameters. The intra-arm and inter-arm off-diagonal connectivity ($V_h$ and $V_v$ respectively) also become function of energy depending upon the intricacy of the structure. This scheme eventually generates an overlap integral $V_d$
between the next nearest neighboring nodes also. Accumulating all these parameters, the difference equation for the effective two-legged ladder can be directly written as follows,
\begin{widetext}
\begin{eqnarray}
(E - \tilde{\epsilon}) \psi_{n,A} & = & V_h (\psi_{n+1,A} + \psi_{n-1,A}) + 
V_v \psi_{n,B} + V_d (\psi_{n+1,B} + \psi_{n-1,B}) \nonumber \\
(E - \tilde{\epsilon}) \psi_{n,B} & = & V_h (\psi_{n+1,B} + \psi_{n-1,B}) + 
V_v \psi_{n,A} + V_d (\psi_{n+1,A} + \psi_{n-1,A}) 
\label{difflad}
\end{eqnarray}
\end{widetext}
%%%%%%%%%%%%%%%%%%%%%%%%%%%%%
The above coupled difference equation is easy to cast in a compact matrix form, viz.,
\begin{widetext}
%%%%%%%%% MATRIX EQUATION %%%%%%%%%%%%%
\begin{eqnarray}
\left [
\left( \begin{array}{cccc}
E & 0 \\
0 & E
\end{array}
\right ) - 
\left( \begin{array}{cccc}
\tilde{u} & V_v \\
V_v & \tilde{u}
\end{array}
\right)
\right ]
\left ( \begin{array}{c}
\psi_{n,A} \\
\psi_{n,B}  
\end{array} \right )
& = & 
\left( \begin{array}{cccc}
V_h & V_d \\ 
V_d & V_h 
\end{array} 
\right)
\left ( \begin{array}{c}
\psi_{n+1,A} \\
\psi_{n+1,B}  
\end{array} \right )
+
\left( \begin{array}{cccc}
V_h & V_d \\ 
V_d & V_h
\end{array}
\right)
\left ( \begin{array}{c}
\psi_{n-1,A} \\
\psi_{n-1,B}
\end{array} \right )
\label{eqladder}
\end{eqnarray}
\end{widetext}
%%%%%%%%%%%%%%%%%%%%%%%%%%%%%%%%%%%%%%%%%
Here the basic strategy is that we can take the advantage of the forms of the potential and hopping matrices, thus formed. It is quite simple to verify that these two matrices do commute with each other and hence they may be simultaneously digonalized. As a consequence of this, one can go for a uniform change of basis and the newly constructed basis is essentially connected to the old one via a similarity transformation, i.e., $f_n = M^{-1} \psi_n$. In the new basis, the two decoupled difference equations for $A$ and $B$ arms respectively will be free from any cross-term and hence they will be linearly independent of each other. The isolated set of equations have the forms, viz.,
\begin{eqnarray}
\left [ E - \left (\tilde{u}+V_v \right ) \right ] 
f_{n,A} & = & (V_h+V_d) ( f_{n+1,A} + f_{n-1,A} ) \nonumber \\
\left [ E - \left (\tilde{u}-V_v \right ) \right ] 
f_{n,B} & = & (V_h-V_d) ( f_{n+1,B} + f_{n-1,B} )
\label{decouple}
\end{eqnarray}
Each of the above set of equations will put their individual mark on the eigenspectrum and the spectral convolution of those two will form the portrait of allowed eigenspectrum of the original geometry. The strength of our analytical attempt lies here. As the mathematical transformation helps us to evaluate the information about the localization of wave packet, the analytical construction of flat band, if any, can also be revealed easily from the above-mentioned discussion. The analysis can be tested also through the numerical evaluation of density of states, transport, band dispersion and other spectral features.

%We now concentrate on the transformed ladder and proceed further.
For the general kagom\'{e} ribbon case, the energy dependent tight-binding parameters of the 
transformed two-leg ladder are listed as follows,
\begin{eqnarray}
\tilde{u} &=& \epsilon_1 + \frac{(E-\epsilon_{1}) (t^2 + t_{1}^2 + \chi_{1}^2)}{\delta} + \frac{2 t_1 \gamma_1 \chi_1}{\delta} \nonumber \\
V_h &=& \frac{(E-\epsilon_{1}) t t_1 + \gamma_1 t \chi_1}{\delta} \nonumber\\
V_d &=& \frac{(E-\epsilon_{1}) t \chi_1 + \gamma_1 t t_1}{\delta} \nonumber\\
V_v &=& \gamma_1 + \frac{\gamma_1 (t^2 + t_{1}^2 + \chi_{1}^2)}{\delta} + \frac{2 t_1 (E-\epsilon_{1}) \chi_1}{\delta}
\label{para1}
\end{eqnarray}
with $\epsilon_1 = \epsilon+ [t^2/(E-\epsilon)]$, $\gamma_1 = t^2/(E-\epsilon) = \chi_1$,
$t_1 = t+ [t^2/(E-\epsilon)]$ and $\delta=(E-\epsilon_1)^2-\gamma_{1}^2$.
These parameters will be helpful in writing the decoupled set of difference equations in the new basis following the Eq.~\eqref{decouple} and these equations can reveal the band diagram for the original kagom\'{e} network. One can also have the idea of the extents of the bands from the decoupled set of equations. 

\subsection{Study of band dispersion}
%%%%%%%%%%%%%%%%%%%%%%%%%%%%%%%%%%%%%%%%%%%%%%%%%%%%%%%%%%%
\begin{figure}[ht]
\centering
\includegraphics[clip,width=4 cm,angle=0]{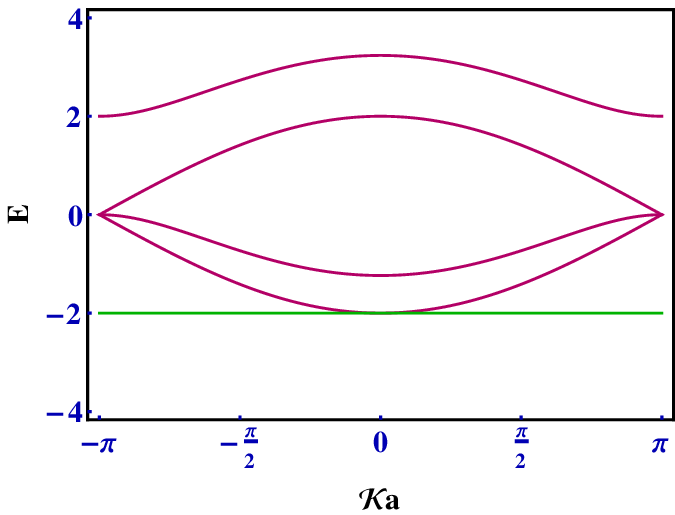}
%(b) \includegraphics[clip,width=5.3 cm,angle=0]{dens3.eps}
%(c) \includegraphics[clip,width=5.3 cm,angle=0]{dens5.eps}\\
%(d) \includegraphics[clip,width=5.3 cm,angle=0]{kagome2trans1.eps}
%(e) \includegraphics[clip,width=5.3 cm,angle=0]{kagome2trans2.eps}
%(f) \includegraphics[clip,width=5.3 cm,angle=0]{kagome2trans3.eps}
\caption{(Color online) Demonstration of band dispersion of the general kagom\'{e} ribbon.
The \textit{flat band} at $E=-2$ is prominent in the plot.}  
\label{disp1}
\end{figure}
%%%%%%%%%%%%%%%%%%%%%%%%%%%%%%%%%%%%%%%%%%%%%%%%%%%%%%%%%%%
%%%%%%%%%%%%%%%%%%%%%%%%%%%%%%%%%%%%%%%%%%%%%%%%%%%%%%%%%%%
\begin{figure*}[ht]
\centering
(a) \includegraphics[clip,width=5.3 cm,angle=0]{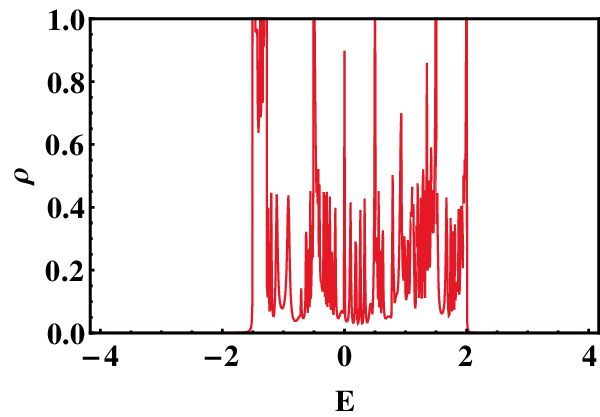}
(b) \includegraphics[clip,width=5.3 cm,angle=0]{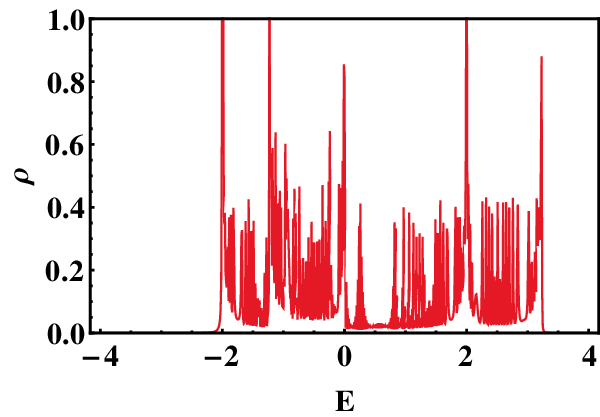}
(c) \includegraphics[clip,width=5.3 cm,angle=0]{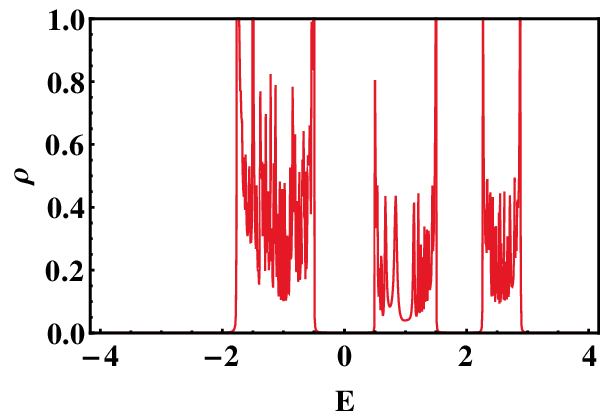}\\
(d) \includegraphics[clip,width=5.3 cm,angle=0]{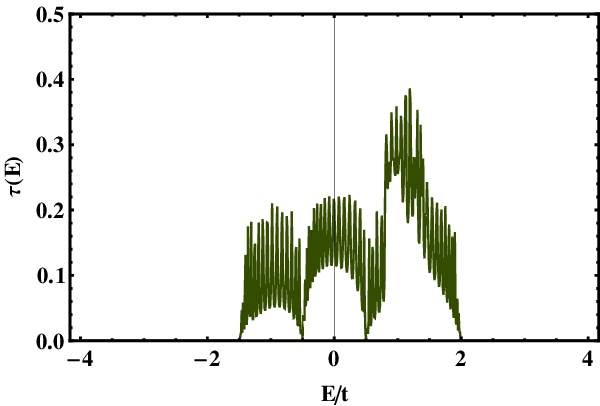}
(e) \includegraphics[clip,width=5.3 cm,angle=0]{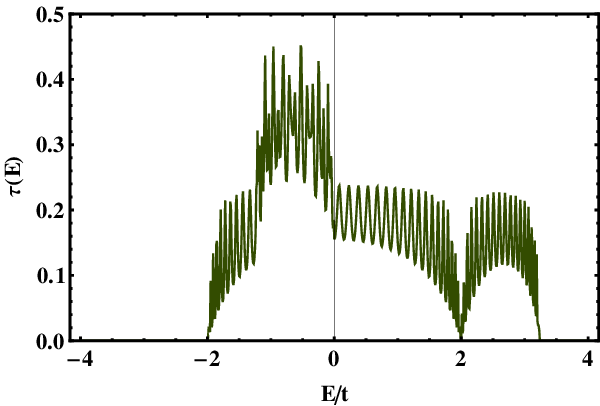}
(f) \includegraphics[clip,width=5.3 cm,angle=0]{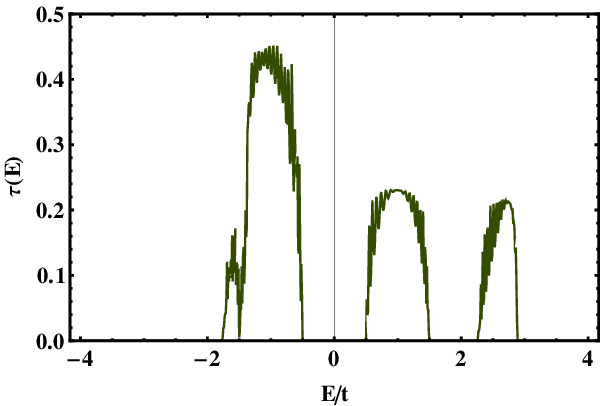}
\caption{(Color online) (Upper panel) Variation of density of states with the energy of the electron
and (lower panel) variation of transmission coefficient with energy for
the general kagom\'{e} ribbon for (a) $x < t$, (b)
$x=t$ and (c) $x > t$ respectively.}  
\label{type2dos}
\end{figure*}
%%%%%%%%%%%%%%%%%%%%%%%%%%%%%%%%%%%%%%%%%%%%%%%%%%%%%%%%%%%
The two linearly independent difference equations respectively demonstrate 
two effective one dimensional chains with the associated tight-binding parameters.
It is then straightforward to evaluate the energy-wave vector relation for those
two tight-binding chains of identical sites in terms of the 
`effective' parameters.
The tight-binding band dispersion relation for the effective 1-d chains ($A$ and $B$
respectively) can be written down as,
\begin{eqnarray}
& E = \epsilon_A + 2 t_A \cos (ka) \nonumber\\
& E = \epsilon_B + 2 t_B \cos (ka)
\label{disp}
\end{eqnarray}
With the help of Eq.~\eqref{decouple} and Eq.~\eqref{para1}, one may easily identify
and evaluate $\epsilon_A, \epsilon_B$ and
$t_A, t_B$ as, $\epsilon_A = \left (\tilde{u}+V_v \right )$, $t_A = (V_h+V_d)$, 
$\epsilon_B = \left (\tilde{u}-V_v \right )$, $t_B = (V_h-V_d)$.
On simplification of the Eq.~\eqref{disp},
the dispersion relations for the two decoupled chains $A$ and $B$ respectively read as,
\begin{eqnarray}
& [E-(\epsilon-2t)] [(E-\epsilon)^{2}-2t \left(E-\epsilon+2t(1+\cos ka)\right)]= 0 \nonumber\\
& (E-\epsilon)^2 - 2 t^2 (1+\cos ka) = 0
\label{simpledisp}
\end{eqnarray}
If we carefully examine the band dispersion relationships, we see that the first part of the first equation of Eq.~\eqref{simpledisp} describes a flat band state indicating the \textit{compact localization} of the single particle eigenstate.
The other $k$-dependent parts show the four dispersive modes, as expected.
It is needless to mention that all the bands exactly match with the band structure obtained from the direct diagonalization of the Hamiltonian matrix, expressed in terms of momentum space. Thus the analytical strategy to unravel the prediction of band structure can be easily applied to any kind of quasi-one dimensional ribbon shaped networks.

\subsection{Evaluation of density of states}
In continuation with the previous analytical proposition of allowed eigenspectrum and the associated study of band dispersion, it is conventional to find out the nature of the density of states (DOS) as a function of the energy of the injected projectile. This will help further to comment on the detailed description of localization aspects offered by this ribbon shaped geometry.
The procedure is the standard green's function formalism which is often used to compute the variation of the number of states per unit energy interval. It is governed by the well-used relation, viz.,
\begin{equation}
\rho=-\left( \frac{1}{\pi} \right) Im [G_{00} (E-H+i \eta)^{-1}]
\end{equation}
$\eta$ is the reasonably small imaginary part added to energy for numerical evaluation.
The results mentioning the intricate look of the spectral density is plotted in the Fig.~\ref{type2dos}
(upper panel). 
The variations are displayed for different values of the relative strengths of the two kinds of hopping parameters, i.e., $x/t$, where, $x$ represents the overlap integral within the unit cell of the kagom\'{e} ribbon and $t$ is the inter-unit cell hopping integral.
We find that for $x/t <1$, the window of allowed eigenmodes is very narrow. As we set
$x = t$, the range of resonant band in the DOS increases 
but no gap is produced resulting in a diffusive band.
%displaying an interesting nature as shown in the Fig.~\ref{type2dos}(b).
 This indicates that with the gradual increment of $x/t$, one can create more number of permissible states. This tendency gets changed as we go for $x/t >1$. A number of subbands with intermediate gaps are observed for such case. Those spectra may offer delocalization or some compact localization depending upon the choice of energy of the electron. This can be validated in the subsequent discussion.

\subsection{Transmission characteristics}
To validate the spectral issues demonstrated, we investigate the variation of transmission characteristics as a function of energy of the incoming electron. This response profile will certainly corroborate our spectral findings. The work out is quite standard and can be done with the aid of basic mechanism of multi-channel transmission. 
This theory can be rigorously applied to several quasi-one dimensional structures.
For this estimation, first we need to fix the underlying network of finite dimension within a pair of semi-infinite periodic leads named as `source' and `drain' respectively. For the lead-system-lead composite network, considering the
 effects of contact electrodes via the self-energy corrections, and with the help of 
 Fisher-Lee relation~\cite{sup,fish} the transmission probability~\cite{paro1}-\cite{paro3} can
 be calculated as,
\begin{equation}
\tau=Tr[\Gamma_S G^r \Gamma_D G^a]
\label{transport}
\end{equation}
where $\Gamma_S$ and $\Gamma_D$ are the coupling matrices,
and $G^r$ and $G^a$
 are the required
 retarded and advanced Green's functions respectively.

In Fig.~\ref{type2dos} (lower panel) we have presented the variation of transmission probability with energy. As we see that the ballistic transmission behavior substantiates the resonant character of the eigenfunctions residing within the continua of the DOS plots. For different relative strength of hopping, the plots are sequentially shown. Change in localization length for different eigenmodes may cause any drastic drop in the transmittance. The lead-to-system coupling is chosen judiciously to minimize the scattering probability at the entrance of the channel. Otherwise, this will degrade the incident amplitude of the excitation. Also the overlap integral for the ordered leads is taken carefully to get the full band picture.

%\subsection{Proposition of general analytical scheme for flat bands}

%\subsection{Band dispersion}

\section{Star-kagom\'{e} ribbon network}
\label{star}
\subsection{Analytical demonstration}
%%%%%%%%%%%%%%%%%%%%%%%%%%%%%%%%%%%%%%%%%%%%%%%%%%%%%%%%%%%
\begin{figure}[ht]
\centering
\includegraphics[clip,width=7.5 cm,angle=0]{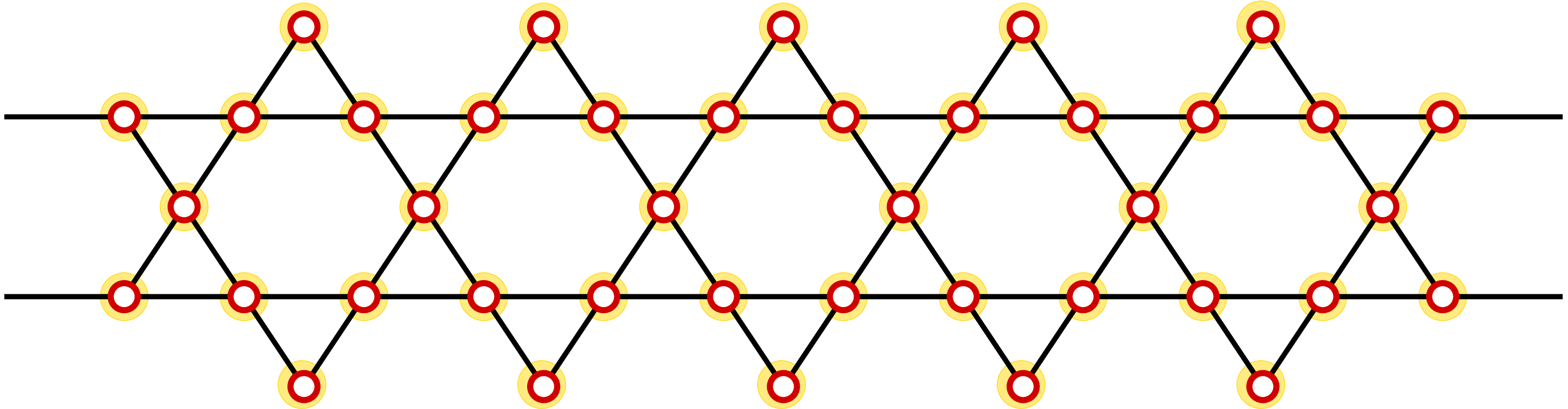}
%(b) \includegraphics[clip,width=5 cm,angle=0]{star-trans1.eps}\\
%(c) \includegraphics[clip,width=5 cm,angle=0]{rgdisp3.eps}
% \includegraphics[clip,width=7.5 cm,angle=0]{SQK-spec4.eps}
\caption{(Color online) A pictorial presentation of 
a section of the quasi-one dimensional star-kagom\'{e} ribbon.}  
\label{lattice2}
\end{figure}
%%%%%%%%%%%%%%%%%%%%%%%%%%%%%%%%%%%%%%%%%%%%%%%%%%%%%%%%%%%
%%%%%%%%%%%%%%%%%%%%%%%%%%%%%%%%%%%%%%%%%%%%%%%%%%%%%%%%%%%
\begin{figure}[ht]
\centering
(a) \includegraphics[clip,width=5 cm,angle=0]{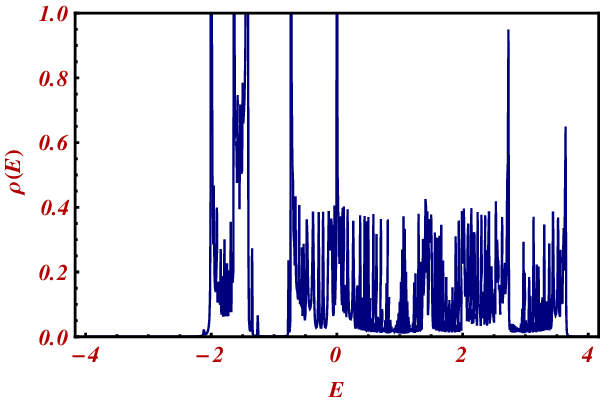}\\
(b) \includegraphics[clip,width=5 cm,angle=0]{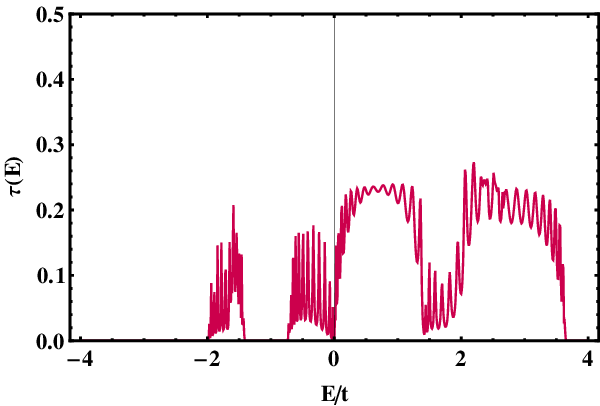}\\
(c) \includegraphics[clip,width=5 cm,angle=0]{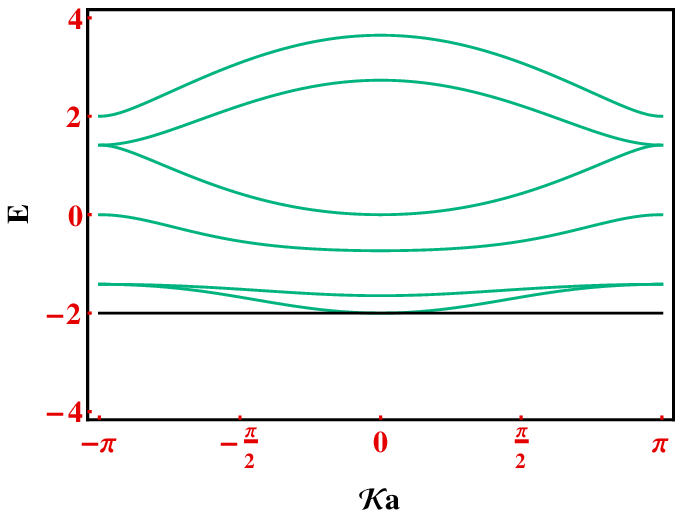}\\
(d) \includegraphics[clip,width=5 cm,angle=0]{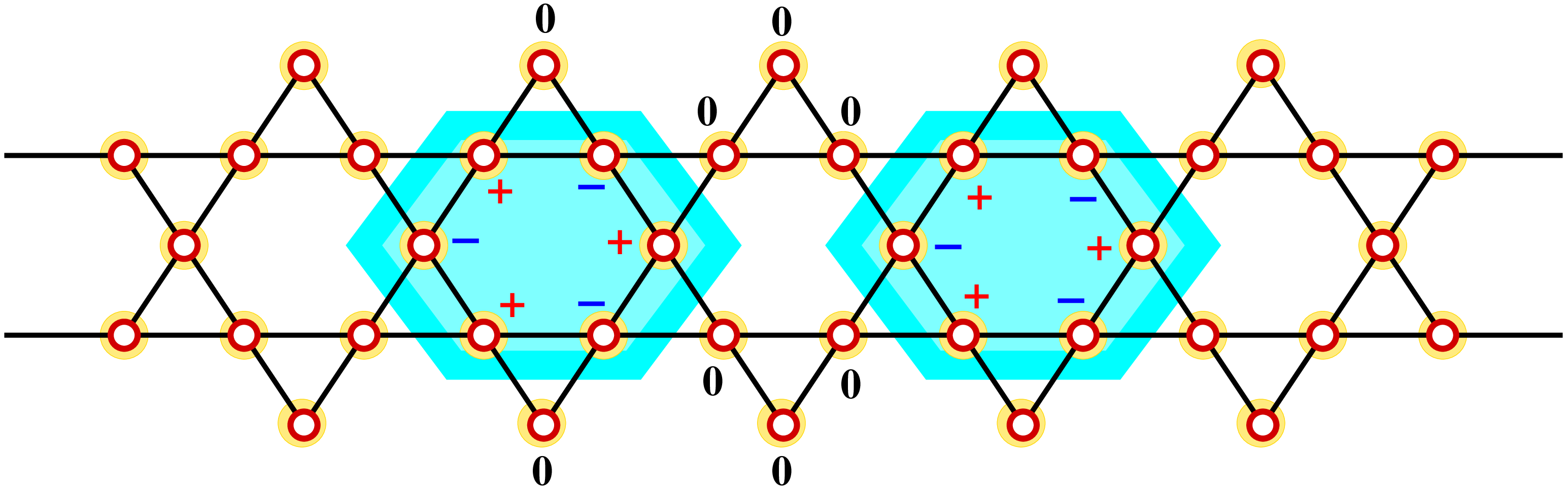}
\caption{(Color online) Demonstration of (a) density of states, (b) transmission probability and 
(c) band dispersion for quasi-one dimensional star ribbon. The flat band at 
$E=-2$ is clearly visible in the dispersion (d) The amplitude distribution for $E=-2$ where the
green shaded regimes are the allowed zones for electron having that energy.}  
\label{starspec}
\end{figure}
%%%%%%%%%%%%%%%%%%%%%%%%%%%%%%%%%%%%%%%%%%%%%%%%%%%%%%%%%%%
The realization of momentum independent flat band(s) in some real materials~\cite{arita}
based on kagom\'{e}-like structures inspires us to apply the unified analytical way for the quasi-one dimensional star ribbon network as shown pictorially in the Fig.~\ref{lattice2}. The basis transformation formalism leads us to work out the \textit{effective} two-legged ladder geometry with corresponding tight-binding parameters. 
The advantage of our scheme is that this analytical way can 
set a uniform route map for
 numerous existing flat band materials~\cite{nyt}-\cite{eye} to explore their transport property and spectral features in a single frame.
The parameters for the decimated ladder are described as follows,
\begin{eqnarray}
\tilde{u} &=& \epsilon_1 + \frac{(E-\epsilon_{1}) (2 t_{1}^2 + \chi_{1}^2)}{\delta} + \frac{2 t_1 \gamma_1 \chi_1}{\delta} \nonumber \\
V_h &=& \frac{(E-\epsilon_{1}) t_{1}^2 + \gamma_1 t_1 \chi_1}{\delta} \nonumber\\
V_d &=& \frac{(E-\epsilon_{1}) t_1 \chi_1 + \gamma_1 t_{1}^2}{\delta} \nonumber\\
V_v &=& \gamma_1 + \frac{\gamma_1 (2 t_{1}^2 + \chi_{1}^2)}{\delta} + \frac{2 t_1 (E-\epsilon_{1}) \chi_1}{\delta}
\label{para2}
\end{eqnarray}
with $\epsilon_1 = \epsilon+ [2 t^2/(E-\epsilon)]$, $\gamma_1 = t^2/(E-\epsilon) = \chi_1$, 
$t_1 = t+ [t^2/(E-\epsilon)]$ and $\delta=(E-\epsilon_1)^2-\gamma_{1}^2$.

It is trivial to write down the linearly independent set of difference equations after decoupling.
 It is to be noted that the localization of incoming excitation is prevailed for any particular energy eigenvalue if it is consistent with both the decoupled difference equations. The localized state (if any) obtained from any of the two equations cannot be distinguished if it falls within the resonant window obtained from the other.

\subsection{Spectral issues}
With the help of the parameters of the decimated ladder one can easily work out the density of states with respect to the energy of the electron as depicted in the Fig.~\ref{starspec}(a). It shows \textit{absolutely continuous} subbands populated by extended type of eigenfunctions. For any energy belonging to the continua the system becomes completely transparent to the injected excitation due to phase coherence.
%The density of eigenstates with energy of the electron is plotted in the Fig.~\ref{starspec}(a). 
The absolutely continuous metallic subbands populated by resonant eigenfunctions is compatible with the periodicity of the lattice. The fine structure of the DOS is related to
 the distribution of spectral weight and is controlled solely by
 the lattice topology. It is important to appreciate the role of the numerical value
 of the imaginary part $\eta$ added to the energy for the numerical workout. It essentially determines
 the spreading of the DOS around an energy. The spectrum supports a compact localized state (CLS) at $E=-2$ which lies at the edge of a subband as observed. The cluster-like amplitude distribution
 (Fig.~\ref{starspec}(d)) for that bound state makes the kinetic signature quenched within the trapping cell.
 
  The variation of transmittivity (Fig.~\ref{starspec}(b)) exhibits the consequent ballistic behavior over the entire allowed range of eigenvalues. The central spiky state cannot show a prominent nature of localization because of its residence inside a conducting subband. Moderate localization length invites an inevitable distinct degradation in the transmission coefficient for that energy as shown. 
  The non-dispersive character of the bound state $(E=-2)$ can be checked from the decoupled set of equations using Eq.~\eqref{decouple}, Eq.~\eqref{disp} and Eq.~\eqref{para2}. The momentum independence
  is validated from the dispersion relation obtained via the decoupled equations (Eq.~\eqref{decouple}).
   Thus the prediction for the presence of CLS becomes true and is pictorially presented in the band dispersion curve (Fig.~\ref{starspec}(c)). It is needless to mention that the dispersion definitely consists of a number of dispersive bands also. The curvature of those bands are different. This indicates an interesting variation in
 the localization lengths corresponding to different modes within the continua and that is reflected in the transmission profile. This is because the curvature $\frac{d^2 E}{dk^2}$ of the $E-k$ diagram is essentially related to the mobility of the wave packet. Thus, the general analytical strategy successfully can demonstrate the intricate nature of the spectral properties offered by this star ribbon.
 
\subsection{Star-kagom\'{e} decorated by fractal geometry}

%%%%%%%%%%%%%%%%%%%%%%%%%%%%%%%%%%%%%%%%%%%%%%%%%%%%%%%%%%%
\begin{figure}[ht]
\centering
(a) \includegraphics[clip,width=4.5 cm,angle=0]{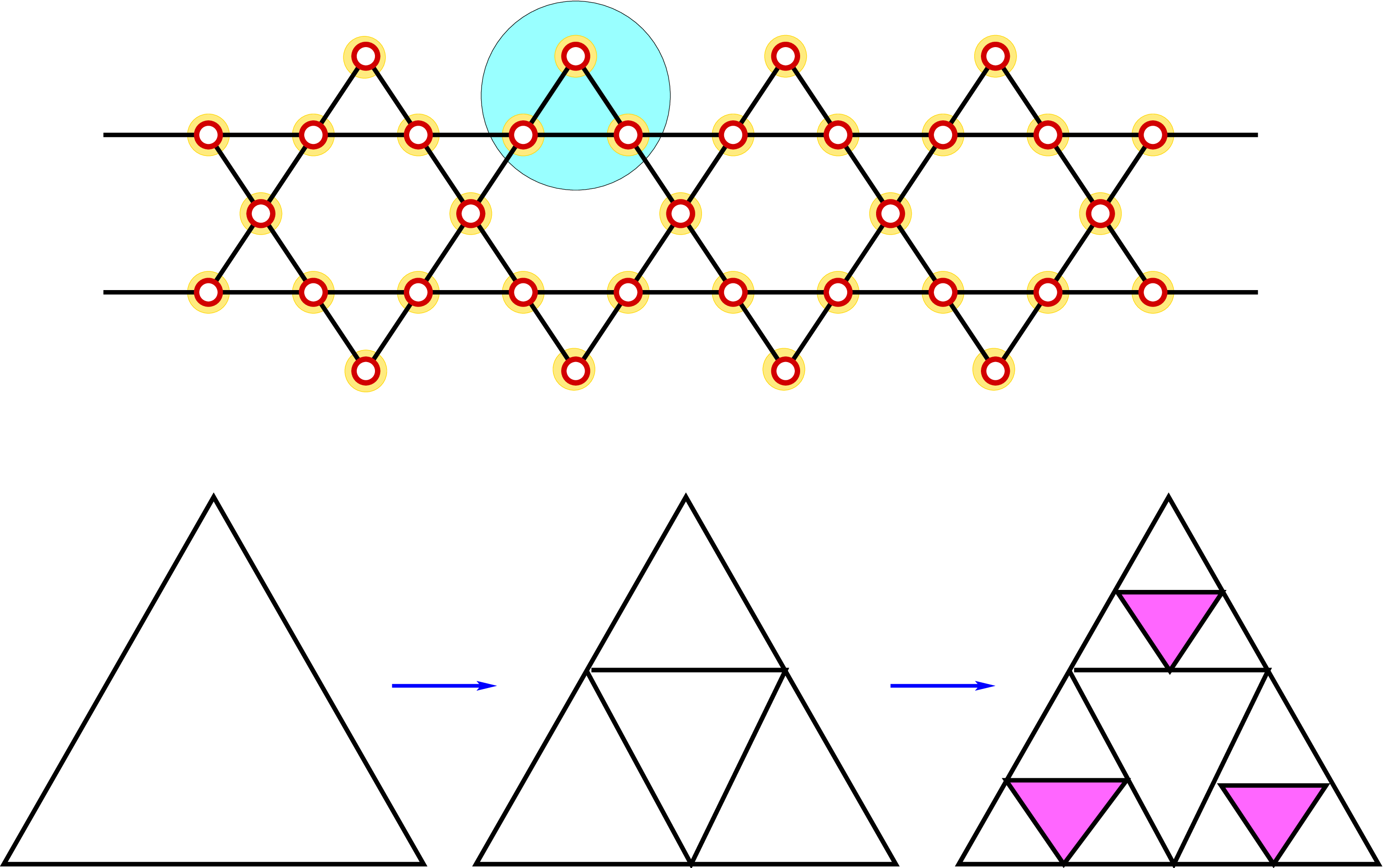}\\
(b) \includegraphics[clip,width=4.5 cm,angle=0]{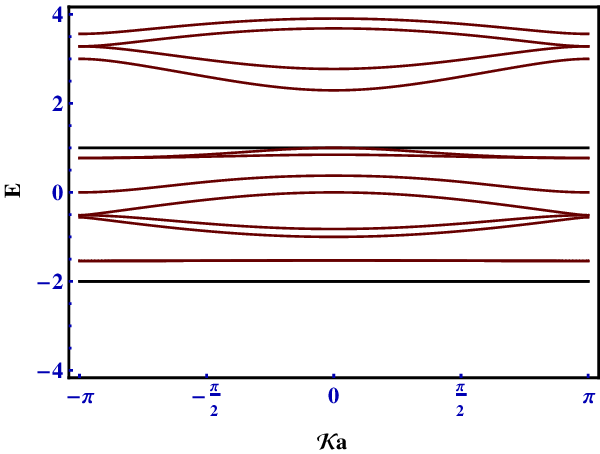}\\
(c) \includegraphics[clip,width=4.5 cm,angle=0]{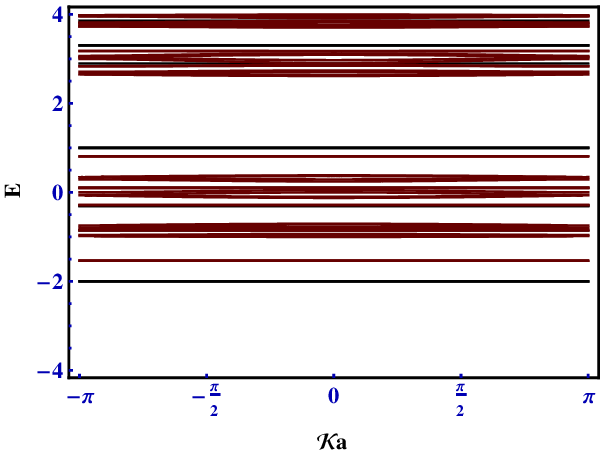}
\caption{(Color online) (a) Schematic view of star
ribbon with a finite generation Sierpinski gasket (SPG) fractal embedded into it.
Plot of dispersion for the above decorated star ribbon with (b) first generation SPG and
(c) third generation SPG.}  
\label{fractalstar}
\end{figure}
%%%%%%%%%%%%%%%%%%%%%%%%%%%%%%%%%%%%%%%%%%%%%%%%%%%%%%%%%%%
In this discussion, we `decorate' the star ribbon by inserting a finite generation Sierpinski gasket (SPG) fractal inside each triangular motif as depicted in the Fig.~\ref{fractalstar}. The unit cell now contains more number of sites depending upon the hierarchy of the fractal. 
The structure with this `decoration' contains a \textit{self-similar} part in its unit cell which
does not carry translational invariance. Although the periodicity is retained globally
in the axial direction. Thus, we may expect an interesting spectral competition which
will produce an exotic band structure.
The general analytical formalism helps us to determine the number of dispersionless flat band modes (for any finite generation of SPG) along with other dispersive bands. It is seen that for any $l$-th generation of the fractal, the band spectrum contains $2^l$ number of flat band (FB) modes which are self-localized by virtue of destructive kind of quantum interference and \textit{local symmetry partitioning}. 
Fig.~\ref{fractalstar} (b) and (c) show the dispersion patterns for finite generation of the fractal. The sequential increase in the number of FB with the generation of the fractal is also presented.
The overall lattice topology plays its part in
mutual cancellation of the phases of the wave function scattered
 off the various lattice sites. For each of such localized states, non-vanishing amplitude profiles are confined within a finite size characteristic trapping island and one such island is effectively isolated from the rest of the lattice by some special vertex with zero wave function amplitude. The areal span of such \textit{confinement cluster} depends certainly on the hierarchy of the fractal.
The off-resonant flat band modes also exhibit an interesting staggering effect in the decay of their associated wave function envelope. By virtue of self-similarity of the fractal one can have the liberty to delay the onset of localization described by those self-localized states with an appropriate choice of scale of length.

Before ending this discussion, we should appreciate that the strength of our analytical attempt is that one can exactly describe the energy-momentum relationship even when unit cell is decorated by a structure having no translational invariance. The scheme is applicable for any finite but large enough generation of the fractal. Thus, when we go for higher generation (close to the thermodynamic limit), the fragmented spectrum will contain large number of flat and dispersive modes situated in a very close proximity of each other forming a quasi-continuous band structure.

\section{Ring-kagom\'{e} ribbon analysis}
\label{ben}
\subsection{Analytical description}
%%%%%%%%%%%%%%%%%%%%%%%%%%%%%%%%%%%%%%%%%%%%%%%%%%%%%%%%%%%
\begin{figure}[ht]
\centering
\includegraphics[clip,width=7.5 cm,angle=0]{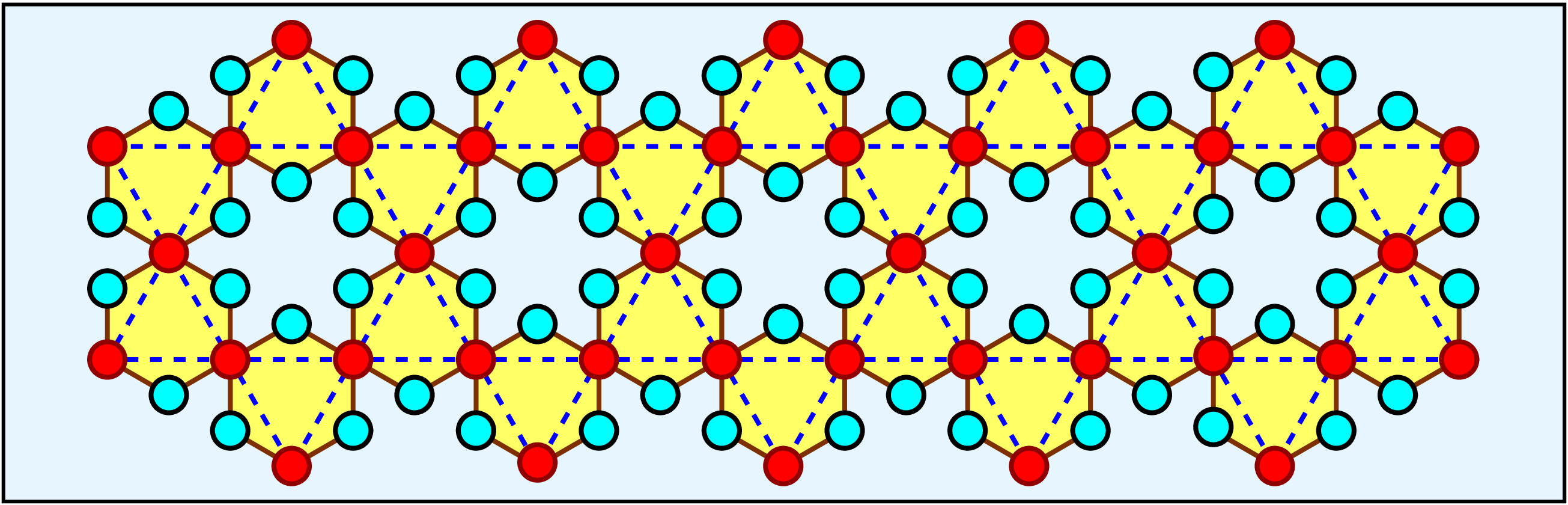}
%(b) \includegraphics[clip,width=5.3 cm,angle=0]{sqkagomedos2.eps}
%(c) \includegraphics[clip,width=5.3 cm,angle=0]{sqkagomedos3.eps}
%(d) \includegraphics[clip,width=3.6 cm,angle=0]{lieb-diados.eps}
\caption{(Color online) A portion of endless ring-shaped kagom\'{e} structure.
The dotted blue lines indicate the decimated star-kagom\'{e} network structure.}  
\label{ring}
\end{figure}
%%%%%%%%%%%%%%%%%%%%%%%%%%%%%%%%%%%%%%%%%%%%%%%%%%%%%%%%%%%
%%%%%%%%%%%%%%%%%%%%%%%%%%%%%%%%%%%%%%%%%%%%%%%%%%%%%%%%%%%
\begin{figure}[ht]
\centering
\includegraphics[clip,width=4.5 cm,angle=0]{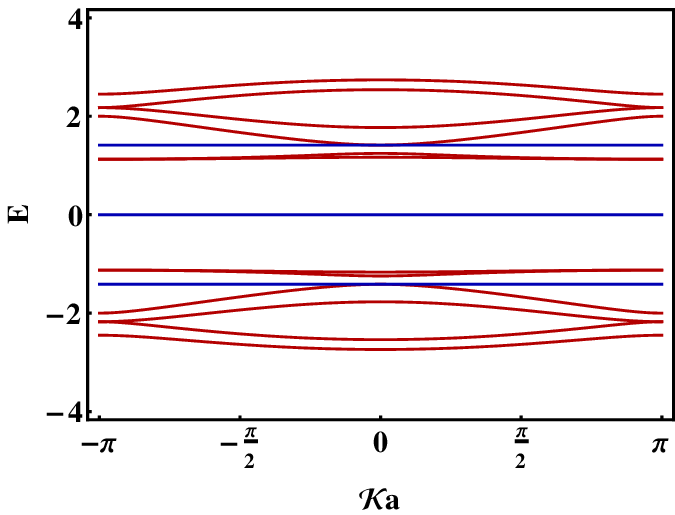}
%(b) \includegraphics[clip,width=5.3 cm,angle=0]{sqkagomedos2.eps}
%(c) \includegraphics[clip,width=5.3 cm,angle=0]{sqkagomedos3.eps}
%(d) \includegraphics[clip,width=3.6 cm,angle=0]{lieb-diados.eps}
\caption{(Color online) Energy-momentum relationship in absence of flux 
for ring shaped kagom\'{e} ribbon.}  
\label{rgdisp4}
\end{figure}
%%%%%%%%%%%%%%%%%%%%%%%%%%%%%%%%%%%%%%%%%%%%%%%%%%%%%%%%%%%
%%%%%%%%%%%%%%%%%%%%%%%%%%%%%%%%%%%%%%%%%%%%%%%%%%%%%%%%%%%
\begin{figure*}[ht]
\centering
(a) \includegraphics[clip,width=5 cm,angle=0]{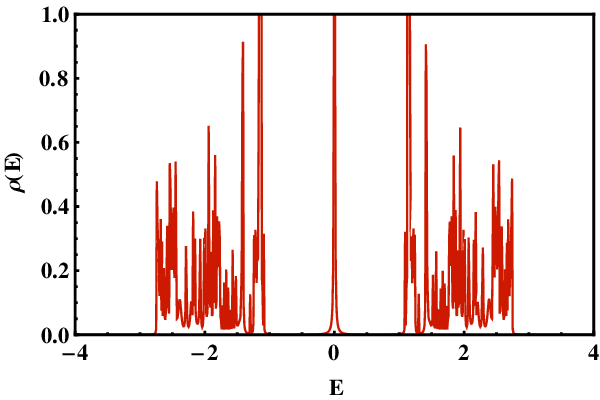}
(b) \includegraphics[clip,width=5 cm,angle=0]{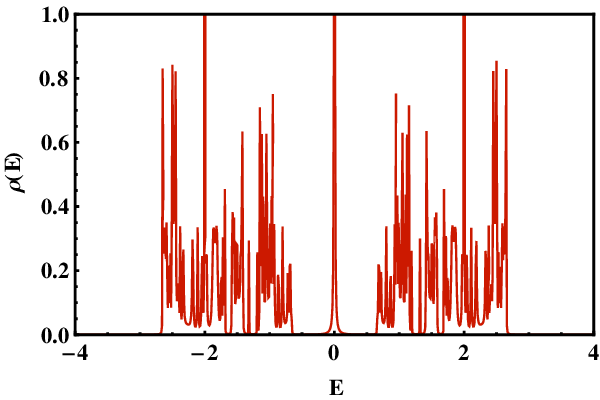}
%(c) \includegraphics[clip,width=3.5 cm,angle=0]{bengenenew1.eps}
(c)\includegraphics[clip,width=5 cm,angle=0]{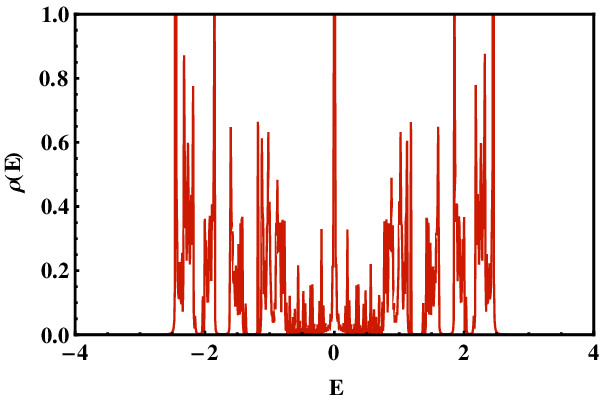}\\
(d) \includegraphics[clip,width=5 cm,angle=0]{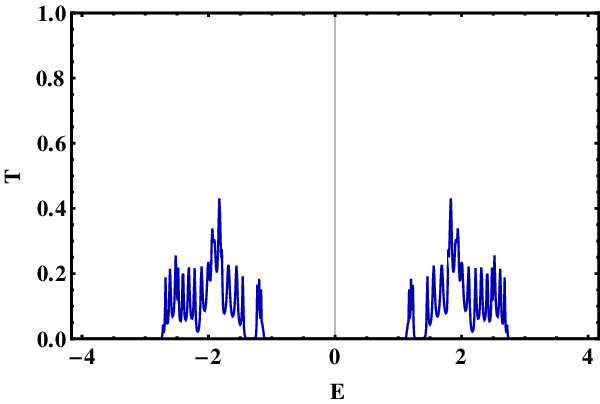}
(e) \includegraphics[clip,width=5 cm,angle=0]{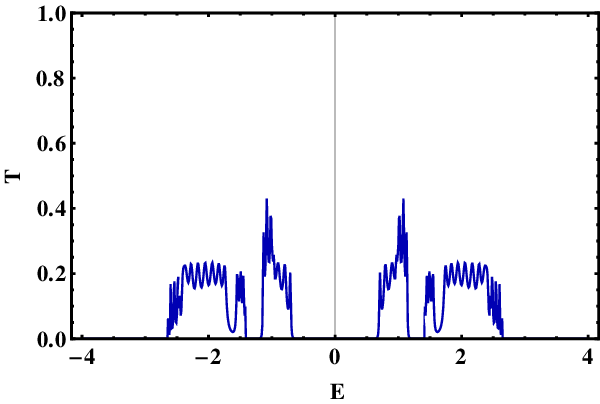}
%(g) \includegraphics[clip,width=3.5 cm,angle=0]{bengenestar-trans1.eps}
(f) \includegraphics[clip,width=5 cm,angle=0]{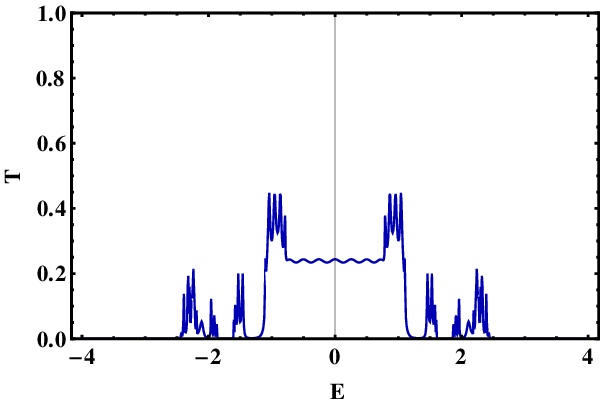}
\caption{(Color online) (Upper panel) Variation of density of states with the energy of the electron 
and (lower panel) that of transmittance with energy for
the benzene shaped kagom\'{e} ribbon with (a) $\Phi=0$, (b)
$\Phi=\Phi_0 /4$ and (c) $\Phi=\Phi_0 /2$ respectively.}  
\label{bengenedos}
\end{figure*}
%%%%%%%%%%%%%%%%%%%%%%%%%%%%%%%%%%%%%%%%%%%%%%%%%%%%%%%%%%%
The next prototype which will be our point of interest in this section is cited in the Fig.~\ref{ring}. Recently, S. Park et al.~\cite{park} reported a van-der-Waals kagom\'{e} lattice material $Pd_{3}P_{2}S_{8}$ which supports compact localized state (CLS) in its band structure resulting in a momentum insensible flat band mode. The CLS is the contribution of all five 4d orbitals of $Pd$, as illustrated. We can follow the same decimation process to map it into an effective star-kagom\'{e} ribbon. The effective parameters of the star ribbon are $\epsilon_2 = \epsilon + 2 t^2/(E-\epsilon)$, $\epsilon_4= \epsilon + 4 t^2/(E-\epsilon)$ and $\tilde{t}=t^2/(E-\epsilon)$. 
$\epsilon_j$ ($j=2$ and $4$) denotes the on-site potential of the site with coordination number $j$.
After that one can move on through identical transformation mechanism as described in the previous section to unravel the spectral property.
The transformed ladder have the parameters as follows, viz.,
\begin{eqnarray}
\tilde{u} &=& \epsilon_1 + \frac{(E-\epsilon_{1}) (t_{1}^2 + t_{2}^2 + \chi_{1}^2)}{\delta} + \frac{2 t_1 \gamma_1 \chi_1}{\delta} \nonumber \\
V_h &=& \frac{(E-\epsilon_{1}) t_1 t_2 + \gamma_1 t_2 \chi_1}{\delta} \nonumber\\
V_d &=& \frac{(E-\epsilon_{1}) t_2 \chi_1 + \gamma_1 t_2 t_1}{\delta} \nonumber\\
V_v &=& \gamma_1 + \frac{\gamma_1 (t_{1}^2 + t_{2}^2 + \chi_{1}^2)}{\delta} + \frac{2 t_1 (E-\epsilon_{1}) \chi_1}{\delta}
\label{para3}
\end{eqnarray}
with $\epsilon_1 = \epsilon_4+ [\tilde{t}^2/(E-\epsilon_2)] + [\tilde{t}^2/(E-\epsilon_4)]$, 
$\gamma_1 = \tilde{t}^2/(E-\epsilon_4) = \chi_1$,
$t_1 = [\tilde{t}^2/(E-\epsilon_4)]$, $t_2 = [\tilde{t}^2/(E-\epsilon_2)]$ and $\delta=(E-\epsilon_1)^2-\gamma_{1}^2$.

The band dispersion for this lattice is shown in the Fig.~\ref{rgdisp4}. 
As observed from the plot that, the band spectrum gets divided symmetrically in two parts with
respect to the band center. The \textit{flat bands} are shown displaying the dispersionless signature.
The corresponding singularity in the DOS is cited later.
The spectrum also supports a number of \textit{quasi-flat} bands whose curvature is extremely small.
This indicates nearly vanishing mobility of the excitation owing to the relatively larger effective mass.
%The central FB is no longer a function of applied magnetic flux. The central gap gradually diminishes with the increase in flux. The curvatures of the dispersive bands are essential function of applied perturbation. Actually, the group velocity of the wave train is controlled by external flux which results in a flux sensitive band curvature modulation.

\subsection{Spectral landscape}
%%%%%%%%%%%%%%%%%%%%%%%%%%%%%%%%%%%%%%%%%%%%%%%%%%%%%%%%%%%
\begin{figure}[ht]
\centering
\includegraphics[clip,width=6 cm,angle=0]{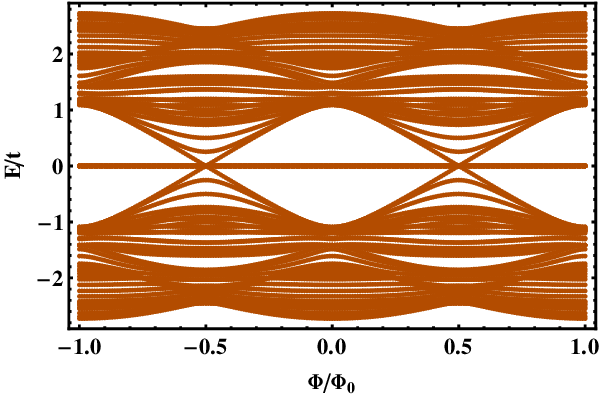}
%(b) \includegraphics[clip,width=5.3 cm,angle=0]{sqkagomedos2.eps}
%(c) \includegraphics[clip,width=5.3 cm,angle=0]{sqkagomedos3.eps}
%(d) \includegraphics[clip,width=3.6 cm,angle=0]{lieb-diados.eps}
\caption{(Color online) Allowed eigenspectrum with respect to external magnetic flux 
for ring shaped kagom\'{e} ribbon.}  
\label{specben}
\end{figure}
%%%%%%%%%%%%%%%%%%%%%%%%%%%%%%%%%%%%%%%%%%%%%%%%%%%%%%%%%%%
The intricate details regarding the variation of DOS with respect to energy for this ring-kagom\'{e}
 ribbon is quite interesting and is demonstrated in the Fig.~\ref{bengenedos} (upper panel). 
The impact of perturbation is also presented in the plots. 
 Here we take each elementary hexagonal plaquette 
 threaded by a uniform magnetic flux $\Phi$. The application of
 a magnetic perturbation breaks the time reversal symmetry of the
 overlap integral along a bond and this is taken care of by
 incorporating a Peierls' phase ($2\pi \Phi/6 \Phi_0$) in the off-diagonal term of the Hamiltonian. As we find that absence of magnetic flux creates a gap at the centre of the spectrum. With the gradual increment of the flux $\Phi (< \Phi_0/2)$ from zero, the gap becomes narrower and at half flux quantum `central gap' closes completely. The central bound state is seen to be present irrespective of any flux value. But at half flux it loses its prominent localization character. This indicates that the specific flux value may invite phase coherence creating some possible quantum path for electron having  such energy and hence the injected excitation diffuses through the network. A complete reverse scenario is seen to happen at the flank of the DOS spectrum. The magnetic flux is observed to create bifurcation in the two main subbands. A number of absolutely conducting subbands separated by mini gaps are viewed in presence of external flux. Thus the external agent of perturbation has a crucial impact on the transport property offered by this quasi-one dimensional ring-kagom\'{e} ribbon. Supportive signature of the response is justified by the plots of the transmission coefficient (Fig.~\ref{bengenedos} (lower panel)) against energy and they are 
 found to be consistent with the DOS profiles, as expected.

In Fig.~\ref{specben}, we have illustrated an overall view of the allowed eigenvalues with respect to the external flux $\Phi$. The central flat band state is robust against the application of perturbation. The energy eigenvalues of the
 underlying network are seen to form mini bands as a function of the
 flux with period equal to $\Phi_0$. There are multiple, inter-twined band
 crossings and quite a dense distribution of eigenvalues, forming
 quasi-continuous $E-\Phi$ band spectrum. In consistent with the previous discussion, it also exhibits a closure of the central gap at $\Phi=\Phi_0/2$ whereas, minigaps are seen at the outer parts of the eigenspectrum. The interesting flux tunable spectrum may throw an achievable challenge to the experimentalists to test our analytical results in the present era of advanced nanotechnology and lithography techniques. Also, encoding the gaps
with appropriate topological quantum numbers remains an
 open problem for this ribbon shaped structure.

\subsection{Aharonov-Bohm oscillation of transmittance}
%%%%%%%%%%%%%%%%%%%%%%%%%%%%%%%%%%%%%%%%%%%%%%%%%%%%%%%%%%%
\begin{figure}[ht]
\centering
(a) \includegraphics[clip,width=5 cm,angle=0]{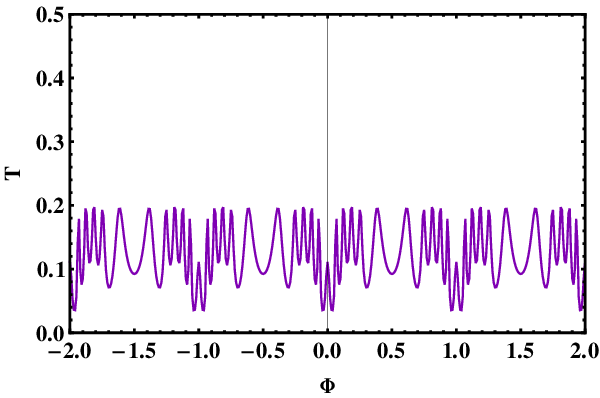}\\
(b) \includegraphics[clip,width=5 cm,angle=0]{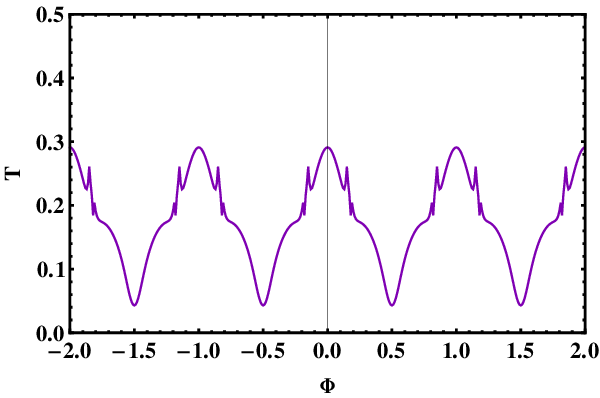}
%(c) \includegraphics[clip,width=6 cm,angle=0]{bengenestar-abosc1.eps}
\caption{(Color online) Aharonov-Bohm oscillation in the
transmission profile with the variation of magnetic flux $\Phi$ for (a) $E=1.45$ and (b)
$E=1.9$ respectively.}  
\label{abosc}
\end{figure}
%%%%%%%%%%%%%%%%%%%%%%%%%%%%%%%%%%%%%%%%%%%%%%%%%%%%%%%%%%%
We have also carefully observed the variation of the transmittance with the applied magnetic flux $\Phi$ for fixed values of the energy of the electron. The transmission coefficient cites a typical oscillatory nature with the variation of the external flux $\Phi$ known as
 the Aharonov-Bohm (AB) oscillation. The patterns, as represented in the Fig.~\ref{abosc}, follows the same flux periodicity consistent with the Fig.~\ref{specben}. We have plotted the AB oscillations in the transmittance sequentially for two different energies, viz., $E=1.45$ and $E=1.9$. It is seen that the localization length corresponding to different eigenmodes ultimately determines the amplitude of the oscillation but the frequency of oscillation is inevitably controlled by the fundamental AB phase factor.
 
\section{Discussion regarding square kagom\'{e} geometry}
\label{sq}
\subsection{Flat band analysis}
%%%%%%%%%%%%%%%%%%%%%%%%%%%%%%%%%%%%%%%%%%%%%%%%%%%%%%%%%%%
\begin{figure}[ht]
\centering
%(a) \includegraphics[clip,width=5.3 cm,angle=0]{sqkagomedos1.eps}
%(b) \includegraphics[clip,width=5.3 cm,angle=0]{sqkagomedos2.eps}
%(c) \includegraphics[clip,width=5.3 cm,angle=0]{sqkagomedos3.eps}\\
%(d) \includegraphics[clip,width=5.3 cm,angle=0]{sqktrans1.eps}
%(e) \includegraphics[clip,width=5.3 cm,angle=0]{sqktrans2.eps}
%(f) \includegraphics[clip,width=5.3 cm,angle=0]{sqktrans3.eps}
\includegraphics[clip,width=7 cm,angle=0]{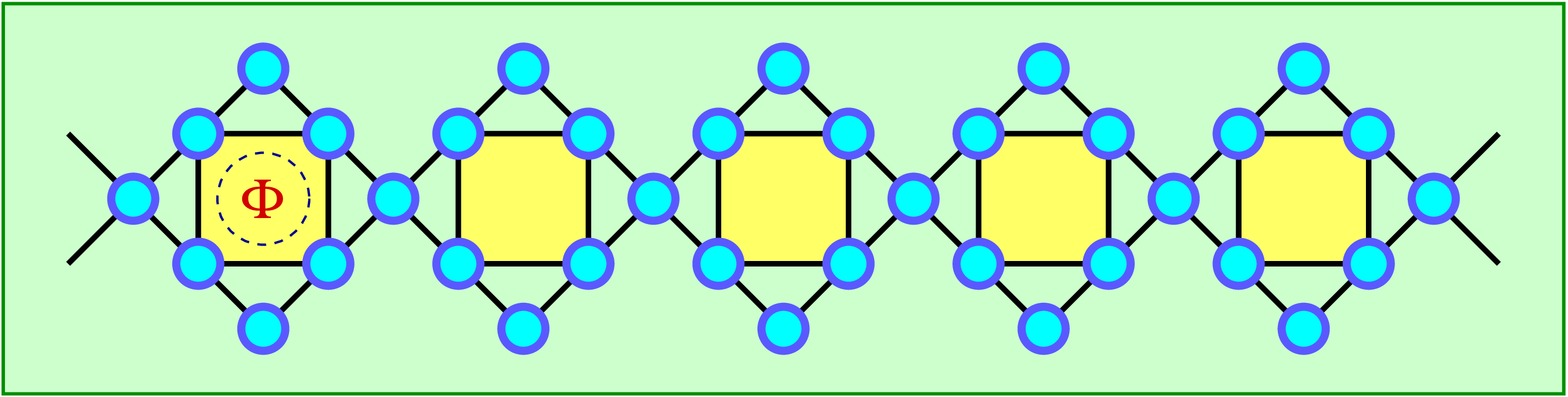}
%(b) \includegraphics[clip,width=5.3 cm,angle=0]{fluxdisp1.eps}
%(c) \includegraphics[clip,width=5.3 cm,angle=0]{fluxdisp2.eps}
%(d) \includegraphics[clip,width=3.6 cm,angle=0]{lieb-diados.eps}
\caption{(Color online) Demonstration of a portion of square kagom\'{e} ribbon with
magnetic flux $\Phi$ embedded into it. }  
\label{lattice4}
\end{figure}
%%%%%%%%%%%%%%%%%%%%%%%%%%%%%%%%%%%%%%%%%%%%%%%%%%%%%%%%%%%

%%%%%%%%%%%%%%%%%%%%%%%%%%%%%%%%%%%%%%%%%%%%%%%%%%%%%%%%%%%
%\begin{figure}[ht]
%\centering
%\includegraphics[clip,width=6 cm,angle=0]{rgdisp1.eps}
%(b) \includegraphics[clip,width=5.3 cm,angle=0]{sqkagomedos2.eps}
%(c) \includegraphics[clip,width=5.3 cm,angle=0]{sqkagomedos3.eps}
%(d) \includegraphics[clip,width=3.6 cm,angle=0]{lieb-diados.eps}
%\caption{(Color online) TO WRITE HERE.}  
%\label{disp1}
%\end{figure}
%%%%%%%%%%%%%%%%%%%%%%%%%%%%%%%%%%%%%%%%%%%%%%%%%%%%%%%%%%%

%%%%%%%%%%%%%%%%%%%%%%%%%%%%%%%%%%%%%%%%%%%%%%%%%%%%%%%%%%%
%\begin{figure}[ht]
%\centering
%(a) \includegraphics[clip,width=7.5 cm,angle=0]{fluxdisp2.eps}\\
%(b) \includegraphics[clip,width=7.5 cm,angle=0]{fluxdisp1.eps}
%(c) \includegraphics[clip,width=7 cm,angle=0]{persistent4.eps}
%(d) \includegraphics[clip,width=2.5 cm,angle=0]{persistent4.eps}
%(e) \includegraphics[clip,width=2.5 cm,angle=0]{persistent5.eps}
%\caption{(Color online) TO WRITE HERE.}  
%\label{fluxdisp}
%\end{figure}
%%%%%%%%%%%%%%%%%%%%%%%%%%%%%%%%%%%%%%%%%%%%%%%%%%%%%%%%%%%

Square kagom\'{e} is one of the prominent prototypes of the kagom\'{e} family. 
A pictorial representation is cited in the Fig.~\ref{lattice4}.
%%%%%%%%%%%%%%%%%%%%%%%%%%%%%%%%%%%%%%%%%%%%%%%%%%%%%%%%%%%
\begin{figure}[ht]
\centering
%(a) \includegraphics[clip,width=5.3 cm,angle=0]{sqkagomedos1.eps}
%(b) \includegraphics[clip,width=5.3 cm,angle=0]{sqkagomedos2.eps}
%(c) \includegraphics[clip,width=5.3 cm,angle=0]{sqkagomedos3.eps}\\
%(d) \includegraphics[clip,width=5.3 cm,angle=0]{sqktrans1.eps}
%(e) \includegraphics[clip,width=5.3 cm,angle=0]{sqktrans2.eps}
%(f) \includegraphics[clip,width=5.3 cm,angle=0]{sqktrans3.eps}
\includegraphics[clip,width=5 cm,angle=0]{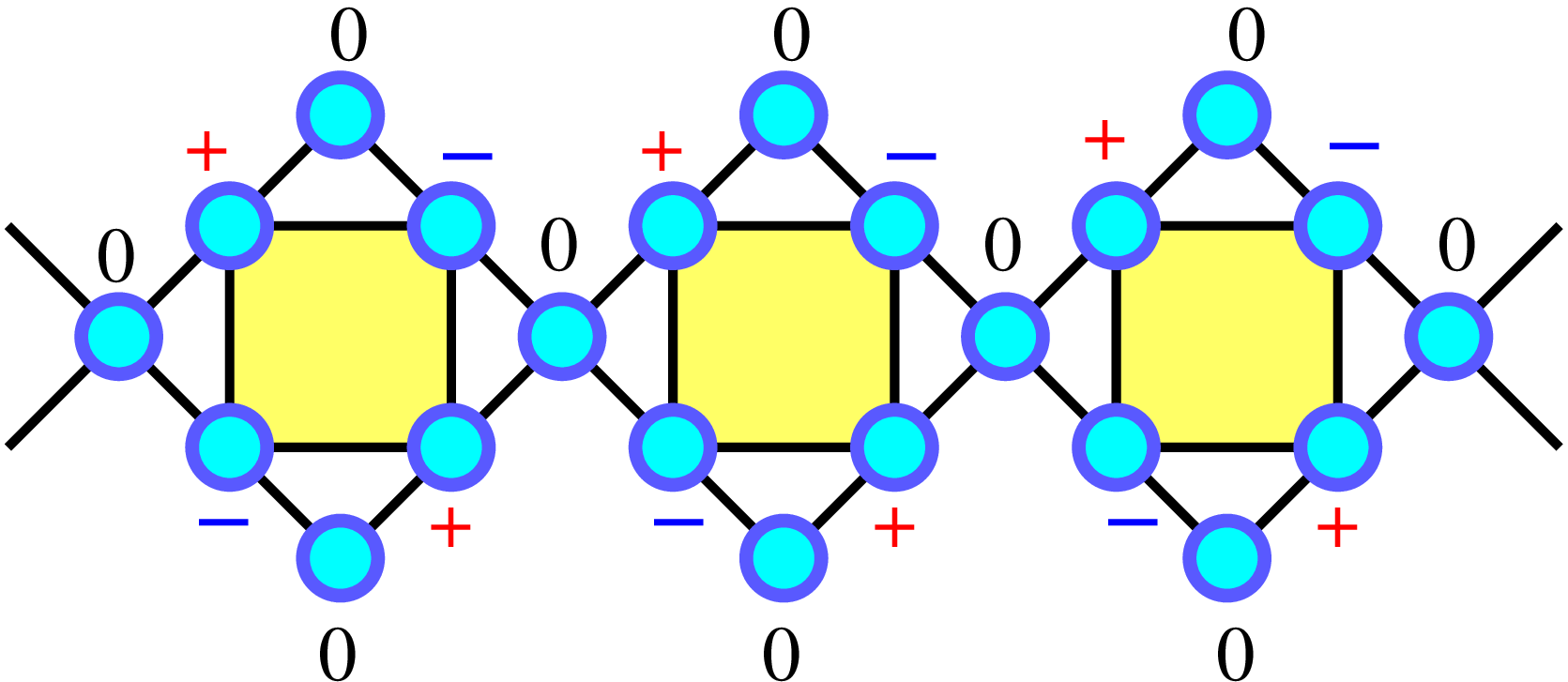}
%(b) \includegraphics[clip,width=5.3 cm,angle=0]{fluxdisp1.eps}
%(c) \includegraphics[clip,width=5.3 cm,angle=0]{fluxdisp2.eps}
%(d) \includegraphics[clip,width=3.6 cm,angle=0]{lieb-diados.eps}
\caption{(Color online) Amplitude configuration displaying the square-shaped 
quantum prison for $E=-2$ (in absence of magnetic perturbation).}  
\label{ampli1}
\end{figure}
%%%%%%%%%%%%%%%%%%%%%%%%%%%%%%%%%%%%%%%%%%%%%%%%%%%%%%%%%%%
The general analytical proposition, as demonstrated in the previous discussion, holds good for this ribbon also. The set of parameters of the decimated quasi-one dimensional ladder are given as follows,
\begin{eqnarray}
\tilde{u} &=& \tilde{\epsilon} + \frac{t_2^2}{\Delta} \left(E-\tilde{\epsilon}\right) 
+ \frac{2t_1^2}{\Delta}\left(E-\tilde{\epsilon} +t_2\right) \nonumber \\
V_h &=& \frac{t_1 t_2}{\Delta}\left(E-\tilde{\epsilon} +t_2\right) \nonumber\\
V_d &=& \frac{t_1 t_2}{\Delta}\left(E-\tilde{\epsilon} +t_2\right) \nonumber\\
V_v &=& t_2 + \frac{t_2^3}{\Delta}+\frac{2t_1^2}{\Delta}\left(E-\tilde{\epsilon} +t_2\right)
\label{para4}
\end{eqnarray}
where, $\tilde{\epsilon}=\epsilon+2 t^2/(E-\epsilon)$, $t_1=t^2/(E-\epsilon)$, 
$t_2=t+t^2/(E-\epsilon)$ and $\Delta=[(E-\tilde{\epsilon})^2 -t_2^2]$.

The next step is quite trivial to write down the linearly independent decoupled set of equations using
Eq.~\eqref{decouple} and Eq.~\eqref{para4}.
It is to be noted from Eq.~\eqref{para4} that the intra-arm hopping $V_h$ and the next nearest neighboring connectivity $V_d$ have the identical energy dependence. 
Hence after the change of basis, the hopping term $(V_h-V_d)$ of the second set of difference equation readily becomes zero, as expected. The solution of $(E - \tilde{u})$ provides the information about the \textit{compact localized state}.
If we set the initial parameters as $\epsilon=0$ and $t=1$ then
the bound states can be exactly calculated as
$E=-2$ and $E=\pm \sqrt{2}$, The associated amplitude distribution for
$E=-2$ is shown pictorially in the Fig.~\ref{ampli1}. Non-zero wave function amplitudes are concentrated inside the \textit{characteristic trapping island} and one such island is effectively `separated' from its neighbors by a vertex at which the amplitude is zero. The frozen kinematics of the incoming electron is therefore the immediate consequence of this cluster-like amplitude profile corresponding to the momentum-free mode ($E=-2$). 
Similar explanation goes for the other two modes as well.
With the help of decimated parameters
the simplified band dispersion relationship turns out to be,
\begin{equation}
(E^3 + 2 E^2-2 E -4)[E^4 -2E^3-6E^2+4(1+E)(1-\cos ka)] = 0
\label{sqkdisp}
\end{equation}
%%%%%%%%%%%%%%%%%%%%%%%%%%%%%%%%%%%%%%%%%%%%%%%%%%%%%%%%%%%
\begin{figure}[ht]
\centering
%(a) \includegraphics[clip,width=5.3 cm,angle=0]{sqkagomedos1.eps}
%(b) \includegraphics[clip,width=5.3 cm,angle=0]{sqkagomedos2.eps}
%(c) \includegraphics[clip,width=5.3 cm,angle=0]{sqkagomedos3.eps}\\
%(d) \includegraphics[clip,width=5.3 cm,angle=0]{sqktrans1.eps}
%(e) \includegraphics[clip,width=5.3 cm,angle=0]{sqktrans2.eps}
%(f) \includegraphics[clip,width=5.3 cm,angle=0]{sqktrans3.eps}
\includegraphics[clip,width=5 cm,angle=0]{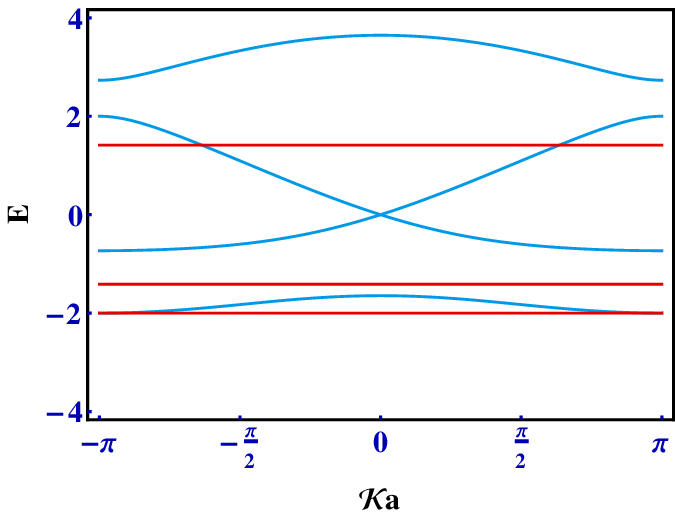}
%(b) \includegraphics[clip,width=5.3 cm,angle=0]{fluxdisp1.eps}
%(c) \includegraphics[clip,width=5.3 cm,angle=0]{fluxdisp2.eps}
%(d) \includegraphics[clip,width=3.6 cm,angle=0]{lieb-diados.eps}
\caption{(Color online) Band dispersion plot in absence of magnetic flux
for the square-kagom\'{e} ribbon geometry. The $k$-independent 
flat bands are seen at $E=-2,\pm \sqrt{2}$ 
(indicated by red lines).}  
\label{sqkdispersion}
\end{figure}
%%%%%%%%%%%%%%%%%%%%%%%%%%%%%%%%%%%%%%%%%%%%%%%%%%%%%%%%%%%
It is clear from the Eq.~\eqref{sqkdisp} that the compact localized states carry the 
non-dispersive signature. The relation is plotted in the Fig.~\ref{sqkdispersion}.

\subsection{Impact of uniform perturbation: flat band engineering}
 %%%%%%%%%%%%%%%%%%%%%%%%%%%%%%%%%%%%%%%%%%%%%%%%%%%%%%%%%%%
\begin{figure}[ht]
\centering
%(a) \includegraphics[clip,width=5.3 cm,angle=0]{sqkagomedos1.eps}
%(b) \includegraphics[clip,width=5.3 cm,angle=0]{sqkagomedos2.eps}
%(c) \includegraphics[clip,width=5.3 cm,angle=0]{sqkagomedos3.eps}\\
%(d) \includegraphics[clip,width=5.3 cm,angle=0]{sqktrans1.eps}
%(e) \includegraphics[clip,width=5.3 cm,angle=0]{sqktrans2.eps}
%(f) \includegraphics[clip,width=5.3 cm,angle=0]{sqktrans3.eps}
\includegraphics[clip,width=5 cm,angle=0]{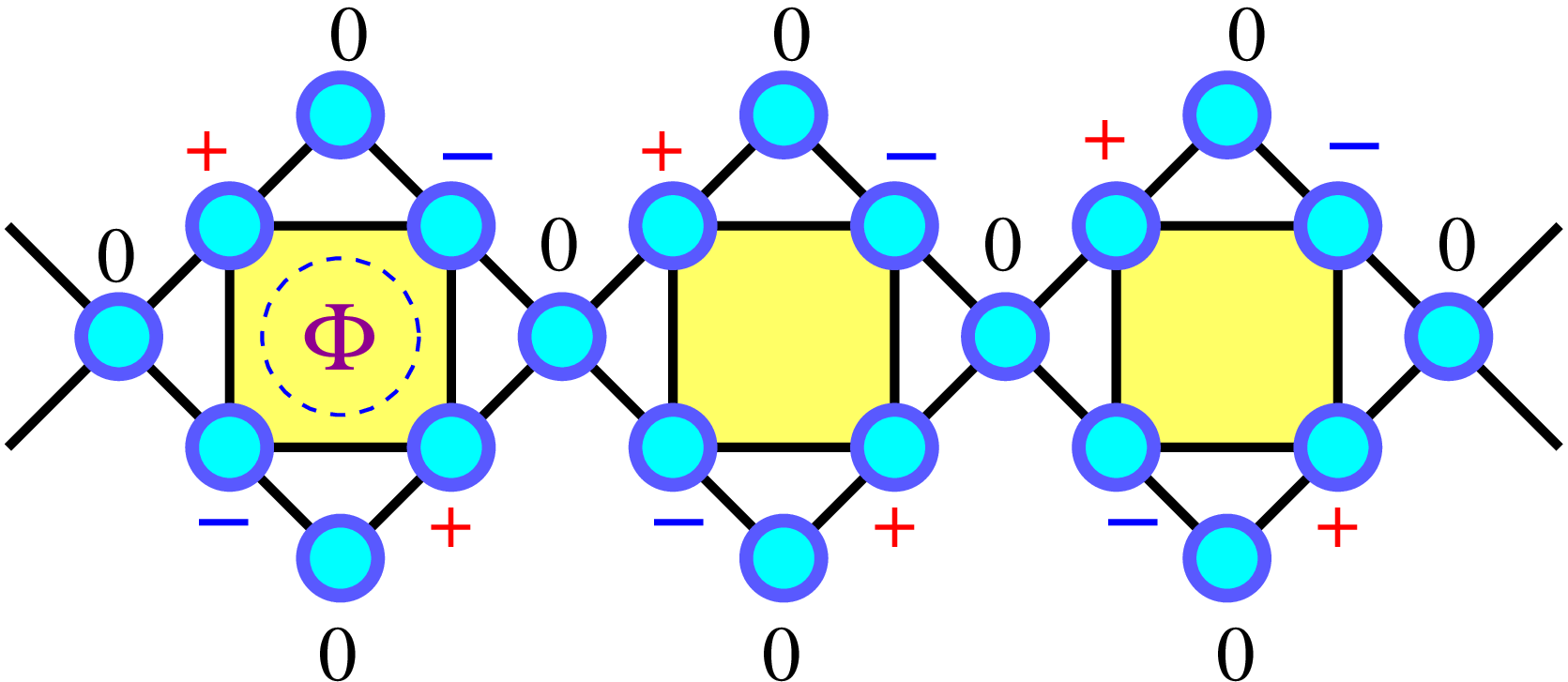}
%(b) \includegraphics[clip,width=5.3 cm,angle=0]{fluxdisp1.eps}
%(c) \includegraphics[clip,width=5.3 cm,angle=0]{fluxdisp2.eps}
%(d) \includegraphics[clip,width=3.6 cm,angle=0]{lieb-diados.eps}
\caption{(Color online) Amplitude distribution corresponding to energy
$E=\epsilon-2 x \cos \Theta$. Here the energy is dependent on external flux $\Phi$.}  
\label{ampli2}
\end{figure}
%%%%%%%%%%%%%%%%%%%%%%%%%%%%%%%%%%%%%%%%%%%%%%%%%%%%%%%%%%%

The application of uniform perturbation essentially invites the standard Aharonov-Bohm phase factor associated with the off-diagonal entries of the Hamiltonian. Here we have considered the square motif as the quantum path of the electron where the kinematics gets attached to the phase factor $\Theta = 2 \pi \Phi/4 \Phi_0$. The choice of the path (i.e., the path along which the 
 Aharonov-Bohm phase factor is calculated) may be justified by an appropriate selection of gauge and it is needless to say that the allowed eigenspectrum and its associated physical aspects are no longer sensitive on that.
 
The magnetic flux invites a non-trivial dependence of allowed eigenspectrum on the external source of perturbation. It is interesting to check that with the help of difference equation, if we set $E=E_0 = \epsilon-2x \cos \Theta$, a consistent solution to the difference equation can be achieved for which the amplitude configuration is cited in the Fig.~\ref{ampli2}. As a consequence of the intermediate connector sites with zero wave function amplitudes, the incoming projectile with that particular energy $E_0$ will loose its mobility and its kinetic energy becomes quenched within the \textit{characteristic trapping island}. Here the interesting fact is that one can have a comprehensive control over the choice of such energy (corresponding to the CLS) with the help of external parameter, i.e, magnetic flux $\Phi$ (in unit of $\Phi_0$).
%%%%%%%%%%%%%%%%%%%%%%%%%%%%%%%%%%%%%%%%%%%%%%%%%%%%%%%%%%%
\begin{figure}[ht]
\centering
\includegraphics[clip,width=6 cm,angle=0]{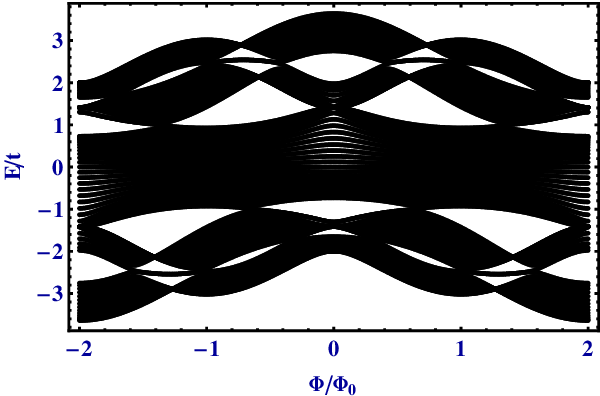}
%(b) \includegraphics[clip,width=5.3 cm,angle=0]{sqkagomedos2.eps}
%(c) \includegraphics[clip,width=5.3 cm,angle=0]{sqkagomedos3.eps}
%(d) \includegraphics[clip,width=3.6 cm,angle=0]{lieb-diados.eps}
\caption{(Color online) Flux sensible periodic eigenspectrum for the square kagom\'{e} ribbon.}  
\label{spec}
\end{figure}
%%%%%%%%%%%%%%%%%%%%%%%%%%%%%%%%%%%%%%%%%%%%%%%%%%%%%%%%%%%

In Fig.~\ref{spec}, we have cited the entire allowed eignespectrum with respect to
magnetic flux $\Phi$. The variation is flux periodic and full of minibands separated by
gaps at the outer part
and at the central part
 of the spectrum as well. Flux controlled range of energy for transmission is easily readable from
 the plot. 
The different curvature of the bands points out the variation of group velocity of the
wave packet with flux. 
 A number of points are seen where the two subbands touch each other resulting a closure
 of gap. Thus one may have a comprehensive manipulation over the general spectral feature provided by
 this square kagom\'{e} ribbon with respect to the application of external perturbation.

\subsection{Discussion of density of states and transport}
%%%%%%%%%%%%%%%%%%%%%%%%%%%%%%%%%%%%%%%%%%%%%%%%%%%%%%%%%%%
\begin{figure*}[ht]
\centering
(a) \includegraphics[clip,width=5.3 cm,angle=0]{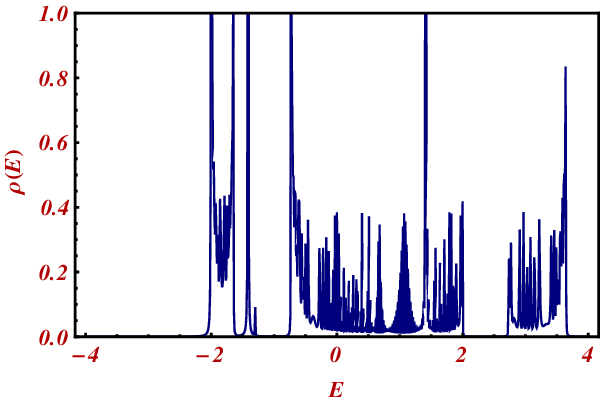}
(b) \includegraphics[clip,width=5.3 cm,angle=0]{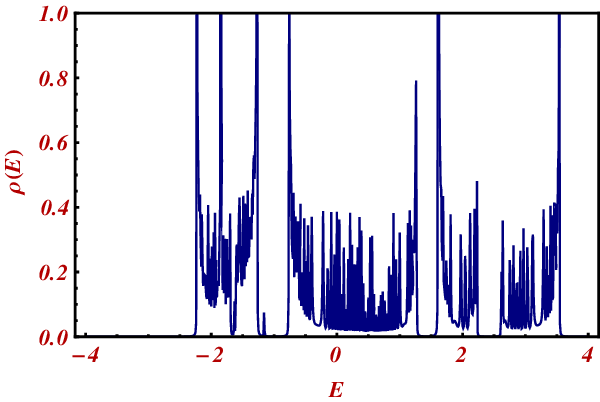}
(c) \includegraphics[clip,width=5.3 cm,angle=0]{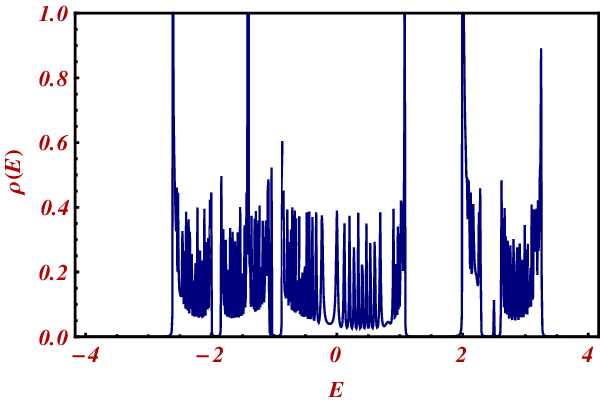}\\
(d) \includegraphics[clip,width=5.3 cm,angle=0]{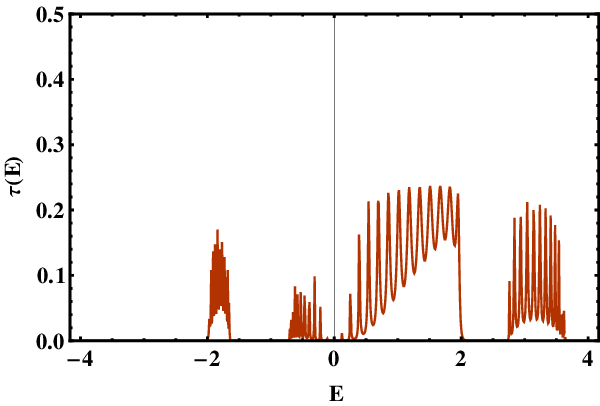}
(e) \includegraphics[clip,width=5.3 cm,angle=0]{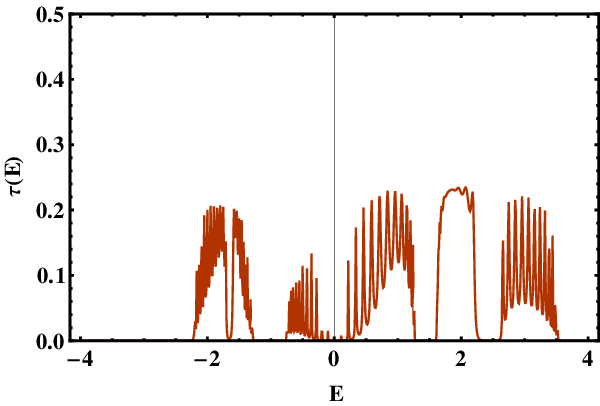}
(f) \includegraphics[clip,width=5.3 cm,angle=0]{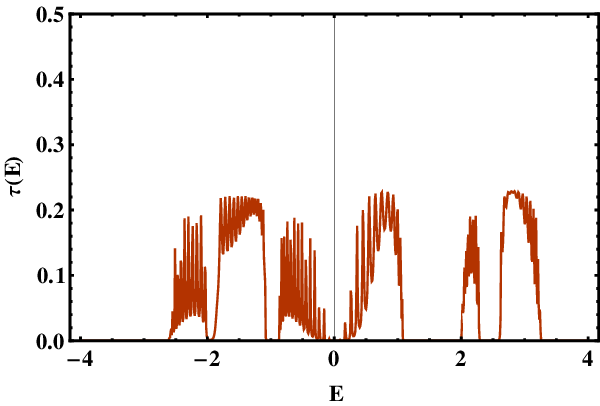}
%(d) \includegraphics[clip,width=5.3 cm,angle=0]{rgdisp1.eps}
%(e) \includegraphics[clip,width=5.3 cm,angle=0]{fluxdisp2.eps}
%(f) \includegraphics[clip,width=5.3 cm,angle=0]{fluxdisp1.eps}
%(d) \includegraphics[clip,width=3.6 cm,angle=0]{lieb-diados.eps}
\caption{(Color online) Graphical variation of density of eigenstates as a function of energy
for the square-kagom\'{e} ribbon geometry with a magnetic flux $\Phi$ embedded into it.
The flux values (in unit of $\Phi_0 = hc/e$) are (a) $\Phi=0$, (b) $\Phi=1/4$ and $\Phi=1/2$ respectively.}  
\label{dos1}
\end{figure*}
%%%%%%%%%%%%%%%%%%%%%%%%%%%%%%%%%%%%%%%%%%%%%%%%%%%%%%%%%%%
In Fig.~\ref{dos1} (a)-(c), we have plotted the variation of density of eigenstates as a function of energy $E$ of the electron for different flux values. As we find that for $\Phi=\Phi_0 /4$, the DOS spectrum becomes subdivided into a number of \textit{absolutely continuous} subbands with gaps in between them. This indicates that with the application of flux, the range of allowed energy becomes highly dependable on the flux $\Phi$. The position of the forbidden gaps may now severely change in accordance with the flux value. The existence of the flux tunable CLS is also prominent in the Fig.\ref{dos1}. As we set $\Phi=\Phi_0 /2$, the subbands slightly adjust their regimes by changing the position of intermediate gaps. However, the central band retains its transparency to the injected electron and therefore does not allow any gap.This flux sensitivity eventually suggests towards a possible mechanism to engineer quantum states with the aid of external source of perturbation in such a ribbon shaped structure. 
Fig.~\ref{dos1} (d)-(f) depicts the transmission behavior for the same ribbon. This 
eventually validates the above 
findings. Continuous subbands in the DOS spectrum expect ballistic transport nature, as observed. 
It is also seen that the resonant windows adjust their positions according to the flux applied.

\subsection{Dispersion in presence of uniform flux}
%%%%%%%%%%%%%%%%%%%%%%%%%%%%%%%%%%%%%%%%%%%%%%%%%%%%%%%%%%%
\begin{figure}[ht]
\centering
%(a) \includegraphics[clip,width=5.3 cm,angle=0]{sqkagomedos1.eps}
%(b) \includegraphics[clip,width=5.3 cm,angle=0]{sqkagomedos2.eps}
%(c) \includegraphics[clip,width=5.3 cm,angle=0]{sqkagomedos3.eps}\\
%(d) \includegraphics[clip,width=5.3 cm,angle=0]{sqktrans1.eps}
%(e) \includegraphics[clip,width=5.3 cm,angle=0]{sqktrans2.eps}
%(f) \includegraphics[clip,width=5.3 cm,angle=0]{sqktrans3.eps}
%(a) \includegraphics[clip,width=5.3 cm,angle=0]{rgdisp1.eps}
(a) \includegraphics[clip,width=5 cm,angle=0]{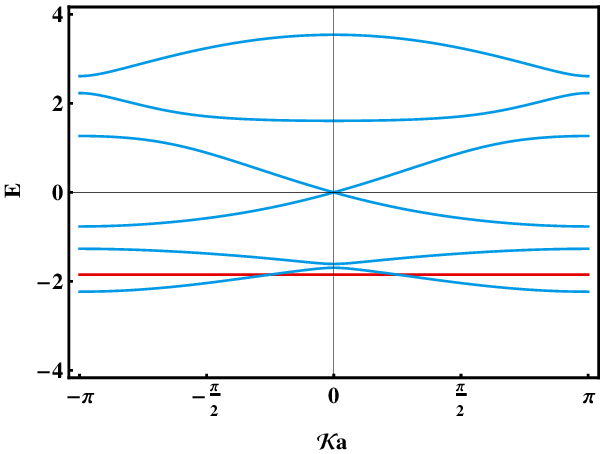}\\
(b) \includegraphics[clip,width=5 cm,angle=0]{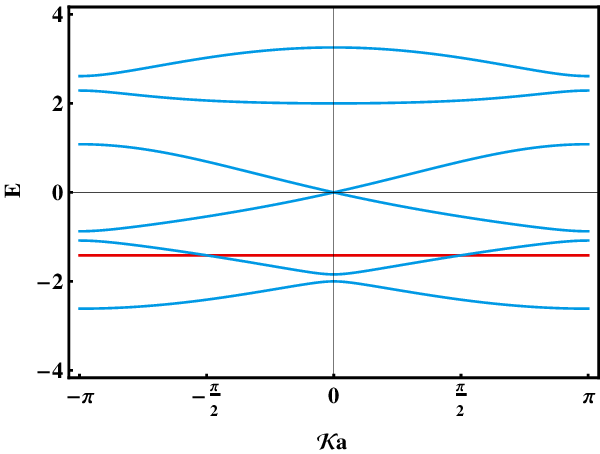}
%(d) \includegraphics[clip,width=3.6 cm,angle=0]{lieb-diados.eps}
\caption{(Color online) Dispersion
for the square kagom\'{e} ribbon geometry with a magnetic flux $\Phi$ embedded into it.
The flux values (in unit of $\Phi_0 = hc/e$) are (a) $\Phi=1/4$ and (b) $\Phi=1/2$ respectively.
The flat line (marked by red color) indicates the flux tunability of the non-dispersive mode.}  
\label{sqkdispersion1}
\end{figure}
%%%%%%%%%%%%%%%%%%%%%%%%%%%%%%%%%%%%%%%%%%%%%%%%%%%%%%%%%%%
For the flux dependent band dispersion, we will utilize the periodicity of the lattice and transform the real space tight-binding Hamiltonian into $k$-space. The transformation relation is quite standard as the basis states are linked with the Wannier states through a Fourier transform. With the help of the recasting formalism, the Hamiltonian for the unit cell (keeping the periodicity in mind) in momentum space reads as,
\begin{equation}
\mathcal{H}(\bm{k})=
\left[ \begin{array}{ccccccc}
\epsilon & t & t & 0 & 0 & t e^{-ika} & t e^{-ika}\\
t & \epsilon & x_f & t & 0 & x_b & 0\\
t & x_b & \epsilon & 0 & t & 0 & x_f\\
0 & t & 0 & \epsilon & 0 & t & 0\\
0 & 0 & t & 0 & \epsilon & 0 & t\\
t e^{ika} & x_f & 0 & t & 0& \epsilon & x_b\\
t e^{ika} & 0 & x_b & 0 & t & x_f & \epsilon
\end{array}
\right]
\end{equation}
with $x_{f(b)}=x e^{\pm i\Theta}$ are the forward and backward hopping integrals
respectively along the square motif 
and $t$ is the hopping along kagom\'{e} structure.
The obvious diagonalization of the matrix leads us to formulate the band dispersion as a function of uniform magnetic flux $\Phi$. Results are displayed in the Fig.~\ref{sqkdispersion1} for quarter and half flux quanta respectively. The plots are appreciably consistent with the DOS profiles, as expected. The flux tunable flat bands at $E=\epsilon -2x \cos \Theta$, i.e., $(E= -1.848)$ for $\Phi=\Phi_0/4$ and $(E= -\sqrt{2})$ for $\Phi=\Phi_0/2$ are the inevitable consequences of the vanishing group velocity of the wave packet
corresponding to such non-diffusive modes.

\subsection{Study of persistent current}
%%%%%%%%%%%%%%%%%%%%%%%%%%%%%%%%%%%%%%%%%%%%%%%%%%%%%%%%%%%
\begin{figure}[ht]
\centering
(a) \includegraphics[clip,width=3.7 cm,angle=0]{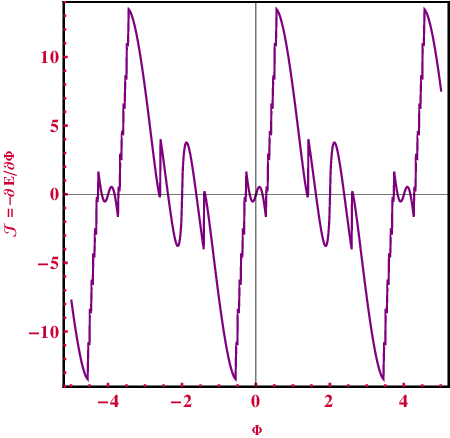}
(b) \includegraphics[clip,width=3.7 cm,angle=0]{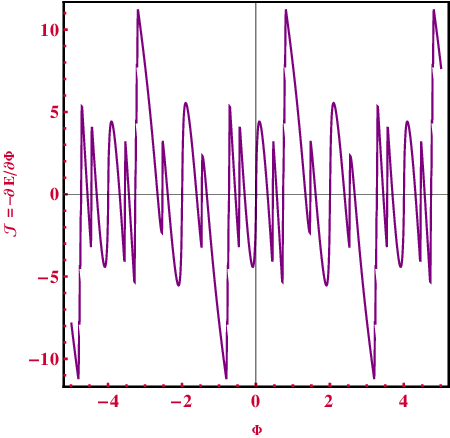}
%(c) \includegraphics[clip,width=5 cm,angle=0]{persistent4.eps}
%(d) \includegraphics[clip,width=2.5 cm,angle=0]{persistent4.eps}
%(e) \includegraphics[clip,width=2.5 cm,angle=0]{persistent5.eps}
\caption{(Color online) Periodic variation of persistent current with flux for (a) $x=1=t$ 
and (b) $x=1, t=0.5$.}  
\label{current}
\end{figure}
%%%%%%%%%%%%%%%%%%%%%%%%%%%%%%%%%%%%%%%%%%%%%%%%%%%%%%%%%%%
For the sake of completeness of the above discussion related to the flux dependent energy spectrum, we may make an estimation of the persistent current~\cite{gefen} which is the dissipationless circulating current  in an isolated mesoscopic ring threaded by magnetic flux. This concept was first introduced by for a single mesoscopic conducting loop structure but later people have investigated the same in multi-loop geometries. This current is certainly a periodic function of the applied perturbation. 
The current as a function of the applied flux $\Phi$ is given by,
\begin{equation}
I_n = -c \frac{\partial E_n}{\partial \Phi}
\end{equation}
This is closely related to the nature of the eigenspectrum and its flux dependence. It is clear from the above equation that the flux modulated change in the curvature of the energy bands should have a direct impression on the behavior the circulating current.
The demonstration of the current for square-kagom\'{e} ribbon is shown in the Fig.~\ref{current} with the external flux. The flux periodic nature is reflected in both the plots. It is also observed that for different relative strengths of hopping parameters, the change in the band curvature in the energy spectrum is eventually reflected in the current profiles. This means when we take $x=1$ and $t=1/2$, more fluctuating behavior in the magnitude is seen. 

\subsection{Aperiodic flux sequence}
%%%%%%%%%%%%%%%%%%%%%%%%%%%%%%%%%%%%%%%%%%%%%%%%%%%%%%%%%%%
\begin{figure*}[ht]
\centering
(a) \includegraphics[clip,width=5 cm,angle=0]{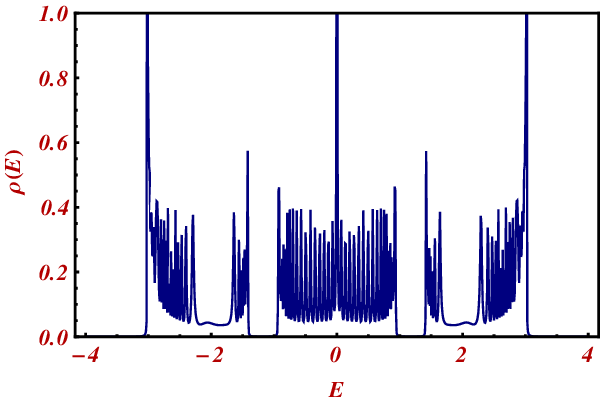}
(b) \includegraphics[clip,width=5 cm,angle=0]{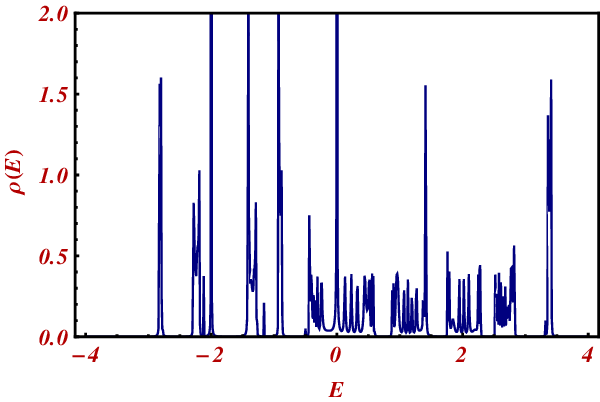}
(c) \includegraphics[clip,width=5 cm,angle=0]{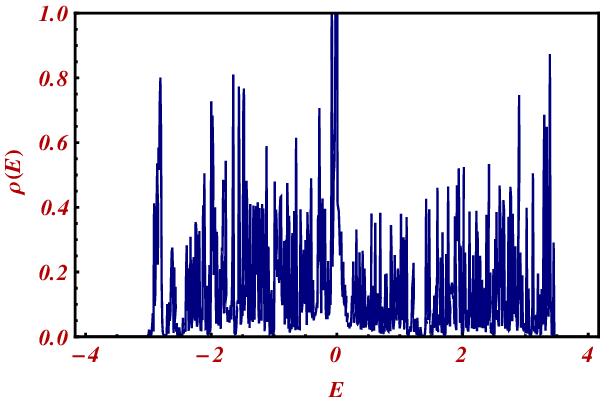}\\
(d) \includegraphics[clip,width=5 cm,angle=0]{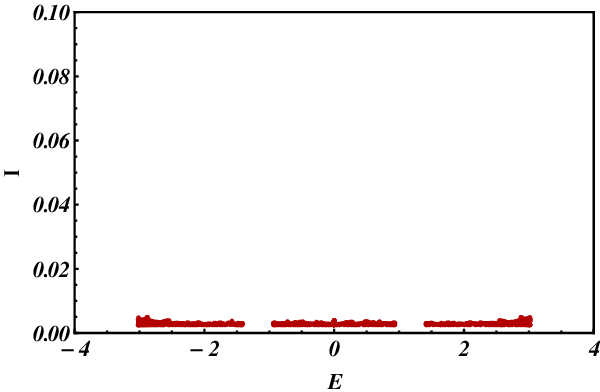}
(e) \includegraphics[clip,width=5 cm,angle=0]{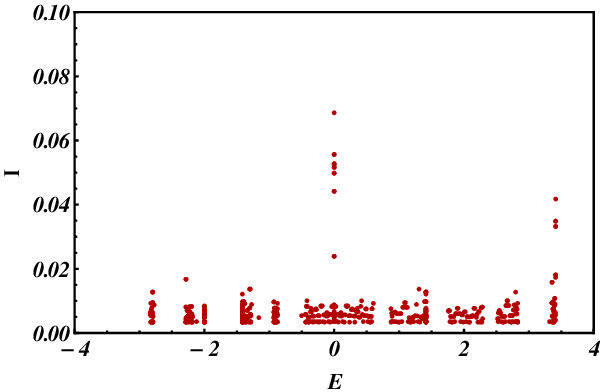}
(f) \includegraphics[clip,width=5 cm,angle=0]{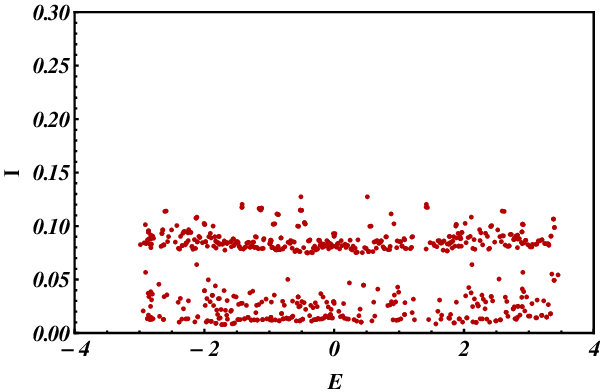}
\caption{(Color online) (Upper panel) Variation of density of states with energy for square kagom\'{e} ribbon and (lower panel) that of inverse participation ratio as a function of energy for the same
with (a) $\Phi_0=1, \beta=1 , \alpha=1$, (b) $\Phi_0=1, \beta= 1/2, \alpha=1$ and (c) $\Phi_0=1, 
\beta= (\sqrt{5}+1)/2, \alpha= 0.8$ respectively.}  
\label{dos2}
\end{figure*}
%%%%%%%%%%%%%%%%%%%%%%%%%%%%%%%%%%%%%%%%%%%%%%%%%%%%%%%%%%%

Now we discuss the same quasi-one dimensional square kagom\'{e} network in tight-binding frame with each square plaquette trapping a magnetic flux that follows a staggered distribution. We have taken a generalization of the Aubry-Andr\'{e}-Harper (AAH) kind of modulation in the magnetic flux pattern. This encodes a non-trivial \textit{effective controllable axial twist} in the original network. The typical flux distribution, for any $j$-th square motif, which incorporates \textit{aperiodic angular twist} can be presented as,
\begin{equation}
\Phi_{j}=\Phi_0 \cos (\pi \beta j^{\alpha})
\label{aah}
\end{equation}
$\Phi_0$ signifies the strength of the applied flux modulation and $\alpha$ denotes the `slowness factor' which has the value in the range between zero and unity. The twisting index $\beta$ controls the frequency of the flux modulation. The two factors simultaneously 
$\beta$ and $\alpha$ introduce the staggering fashion in the magnetic flux profile.
The lower limit of $\alpha$, i.e., $\alpha=0$ demonstrates a constant flux for a given $\beta$, and the upper limit of $\alpha=1$ brings back the AAH variation in the same. It was shown that a non vanishing $\alpha$ can trigger an
interesting 
insulator-metal transition and one could analyze the existence of mobility edges~\cite{das2}, a result which is not found in the conventional AAH model. To the best of our knowledge, the impact of application of such an interesting modulation in the variation of magnetic flux trapped in a quasi-one dimensional kagom\'{e} ribbon has not been raised before.

The selection of such modulation factor in the distribution of magnetic perturbation introduces a generalized flux variation in the square plaquettes. For instance, with $\alpha=1$, one may choose $\beta$ as $1,1/3,$ or any irrational number (golden ratio $\tau = \frac{(\sqrt{5}+1)}{2})$ to incorporate uniform, staggered or quasiperiodic modulation of flux, respectively.

The variation of DOS with energy is shown in the Fig.~\ref{dos2} for the square kagom\'{e} geometry with a deterministic flux modulation. When we set $\beta=\alpha=1$, i.e., a constant flux appears in every square plaquette. As a result we see the resonant subbands separated by intermediate gaps. 
A central spike carrying an apparent signature of localization is also viewed. However, the mode cannot exhibit its
non-diffusive behavior as it lies within the absolutely continuous band.
As we apply staggered magnetic perturbation to the system,
the spectrum shows distinct fragmentation by creating minigaps in between the subbands. The 
width of the individual regime of resonant conduction becomes narrower. 
The the central spike ($E=0$) is still present and is found to robust against any kind of 
flux modulation.
Scenario is changed drastically as one goes for aperiodic limit of the frequency and fractional slowness factor, describing a gapless spectrum. The spectral change is definitely governed by the interplay of the parameters $\beta$ and $\alpha$. The spectral issue is validated with the supportive evaluation of 
\textit{inverse participation ratio} (IPR) as shown in the Fig.~\ref{dos2} (lower panel).
As IPR is a measure of localization length, the spectral modification due to the variation
of the parameters related to the flux, is directly reflected in the IPR plots. Whereas, for the first
case, nearly vanishing IPR indicates extended signature of the states belonging to continua, but for the
aperiodic limit, finite IPR demands the existence of some states having moderate localization lengths.

\subsection{Creation of quantum butterfly}
%%%%%%%%%%%%%%%%%%%%%%%%%%%%%%%%%%%%%%%%%%%%%%%%%%%%%%%%%%%
\begin{figure*}[ht]
\centering
(a) \includegraphics[clip,width=6 cm,angle=0]{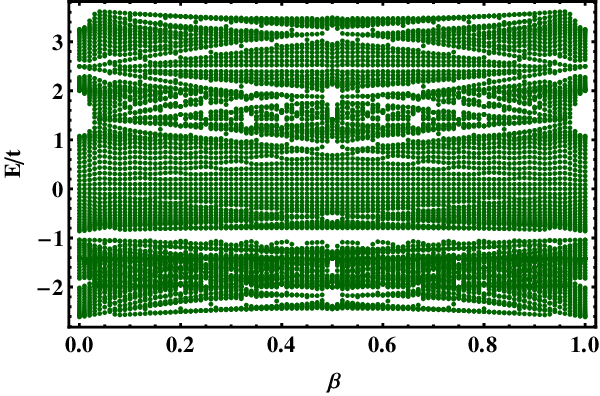}
(b) \includegraphics[clip,width=6 cm,angle=0]{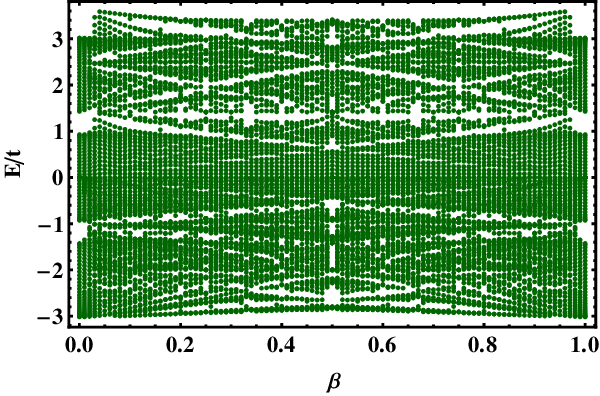}\\
(c) \includegraphics[clip,width=6 cm,angle=0]{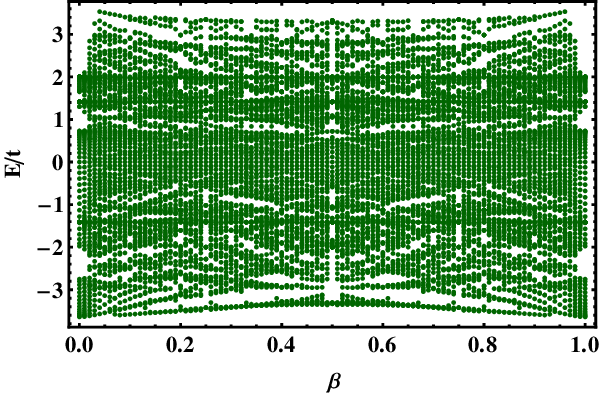}
(d) \includegraphics[clip,width=6 cm,angle=0]{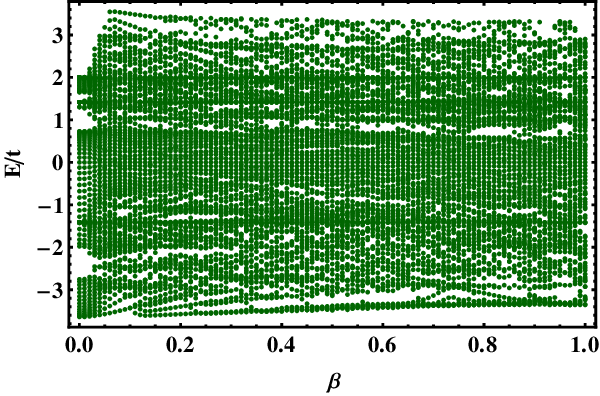}
\caption{(Color online) Energy landscape for square kagom\'{e} ribbon with
(a) $\Phi_0=1/2$, $\alpha=1$, (b) $\Phi_0=1$, $\alpha=1$, (c)
$\Phi_0=2$, $\alpha=1$, and (d) $\Phi_0=2$, $\alpha=0.8$.}  
\label{butter}
\end{figure*}
%%%%%%%%%%%%%%%%%%%%%%%%%%%%%%%%%%%%%%%%%%%%%%%%%%%%%%%%%%%
Before going to the detailed analysis, it is necessary to acquire the information regarding the energy spectrum in respect of such flux distribution. To demonstrate, in this section we present the spectral landscape of the flux controlled square kagom\'{e} network as a function of the modulation frequency $\beta$ for different strengths of the flux. As it is seen in the case of standard AAH model, here also we observe that a variation in $\beta$ creates quantum butterflies in the energy landscape. The interplay of the two parameters $\Phi_0$ and $\beta$ is solely responsible for the butterfly pattern. We utilize the straightforward diagonalization of the Hamiltonian for a system size $N=141$ to obtain the eigenspectrum.

To observe the impact of aperiodic flux modulation, we start from a low strength of the perturbation, i.e., $\Phi_0 = 1/2$ keeping the slowness factor $\alpha=1$. We see an overall uniform energy landscape, the spectrum looks quasi-continuous along with the existence of gaps. A notable change
 in the energy spectrum is observed when we tune $\Phi_0 = 2$. Each of
the Landau levels starts fragmentation producing smaller subbands and ultimately a
 self-similar pattern in the energy portrait. In Fig.~\ref{butter}(c) creation of subbands is prominent as one goes away from the center. This is identical to the Hofstadter butterfly
 and can be viewed as one of its interesting variant. 
Our results
indicate that a quasi-one dimensional network which may be approximated as a
strip geometry of the Hofstadter model is capable of creating a self-similar landscape. The pattern gets destroyed when we turn on the slowness factor $\alpha$. It has also been checked that any fractional value of $\alpha$ cannot produce quantum butterfly.

\section{Extension to photonics}
\label{photo}
%%%%%%%%%%%%%%%%%%%%%%%%%%%%%%%%%%%%%%%%%%%%%%%%%%%%%%%%%%%
\begin{figure}[ht]
\centering
(a) \includegraphics[clip,width=4.5 cm,angle=0]{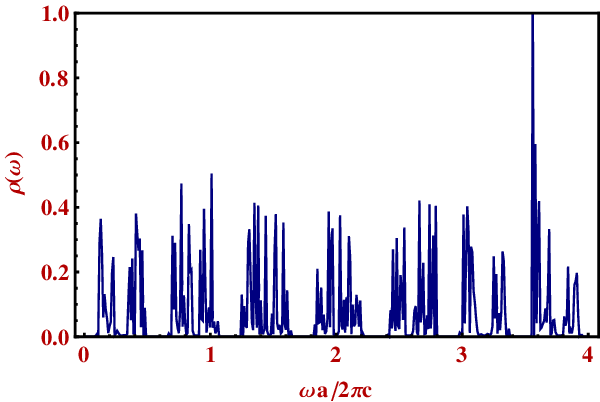}\\
(b) \includegraphics[clip,width=4 cm,angle=0]{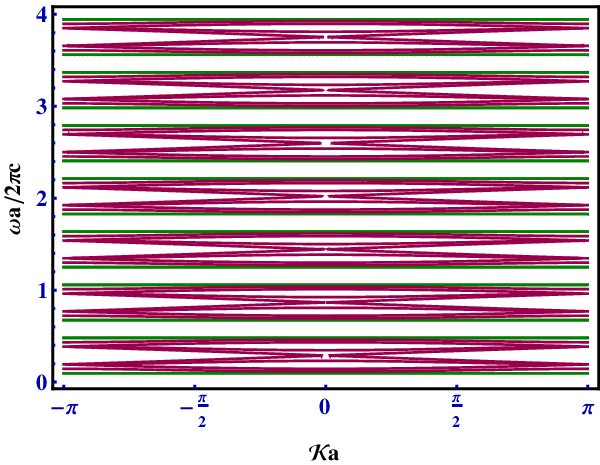}
%(c) \includegraphics[clip,width=7.5 cm,angle=0]{SQK-spec3.eps}
%(d) \includegraphics[clip,width=7.5 cm,angle=0]{SQK-spec4.eps}
\caption{(Color online) (a) Variation of optical density of eigenmodes with frequency and (b)
Photonic dispersion relation for star kagom\'{e} network.}  
\label{opt}
\end{figure}
%%%%%%%%%%%%%%%%%%%%%%%%%%%%%%%%%%%%%%%%%%%%%%%%%%%%%%%%%%%
Localization of classical wave in different medium has been under intensive study with a straightforward analogy with the localization of electron. Photonic band gap (PBG) materials in this context have remained an active topic of research in condensed matter physics over the past few years~\cite{john,ya}. 
The existence of disallowed regimes in photonic crystals where electromagnetic wave are forbidden has immense implications in both fundamental science and technological applications.
Hence with such networks, we come across the interesting non-trivial possibility of manipulating the propagation of an electromagnetic excitation by creating gaps in the band structure of synthetic periodic dielectric structures~\cite{ms}. All these studies constitute an interesting part of mesoscopic physics, which focus on the possibility of localization of light~\cite{zh,call}, a prominent example of which has already been reported for a strongly scattering medium of semiconductor powders~\cite{ad}.

The analytical description provided for the localization of electron has a close parallelism with the optical scenario. This interesting and pertinent issue was initially discusses by 
Sheng et al.~\cite{alex,sheng}, where they presented
 that the propagation of electromagnetic wave through any quasi-one dimensional monomode wave guide network has a
 direct correspondence with the same electronic model with proper initial conditions. The analogy is quite straightforward within the tight-binding formalism. This one-to-one parallelism is purely a mathematical construction and once that is done the
 decimation scheme  becomes insensitive to whether the input appears from a
 quantum background or a classical one.

Following Sheng et al., we can think of a single channel wave guide model constructed by the segments having
 the same dimensions arranged in the form of a star ribbon. We can mention that the choice of dielectric parameter of the core material of the wave guide can set the range of frequency to be propagated within it. In Fig.~\ref{opt}, we have shown the variation of density of photonic modes with frequency and the $\omega-k$ photonic dispersion relation within the range $0<\omega<4$. The refractive index of the core is set as $\mu = \sqrt{3}$ for numerical evaluation. From the plots, it appears that modes are distributed over the entire range and the minibands are separated by gaps. The dispersion also shows several flat, non-dispersive photonic modes which are localized essentially due to the destructive quantum interference. Thus the proposed wave guide structure may act as a suitable prototype for photonic bandgap (PBG)~\cite{ya} kind of 
system. Further, this analogical treatment can also be done for other kagom\'{e} prototypes as well.
%%%%%%

\section{Summary}
\label{closing}
In conclusion, we find that a unified analytical strategy provides a common platform to discuss the spectral issues of different kinds of quasi-one dimensional kagom\'{e} prototypes. Real space decimation formalism helps us to demonstrate the localization of single particle eigenstates, flat band engineering, persistent current study within the tight-binding framework. The discussion is supported by the numerical evaluation of density of states, transmission characteristics and band dispersion and other related physical quantities. We observe that for general kagom\'{e} ribbon, there is a severe change in the spectrum with the variation of relative strength of hopping parameter. The flat band is also worked out by virtue of the basis transformation technique discussed. 
Moreover, for star kagom\'{e} ribbon, spectral property is discussed elaborately following
a in-depth knowledge of transmission profile. The
non-trivial hierarchical distribution of \textit{flat band} states is also illustrated when the unit cell is decorated by a self-similar fractal structure. For ring kagom\'{e} ribbon magnetic flux controlled band tuning is studied in details. This is corroborated with the demonstration of Aharonov-Bohm oscillation of the transmittance with flux. In the case of square kagom\'{e} type, spectral property in presence of uniform flux is reported with the study of persistent current. Further, it has been shown that a deterministic aperiodic variation in the flux can create quantum butterfly in the energy landscape of such prototype.
An interesting analogous extension to photonics is also mentioned for application purpose.

%%%%%%%%%%%%%%%%%%%%%%%%%%%%%%%%%%%%%%%%%%%%%%%%%%%%%%%%%%%

%%%%%%%%%%%%%%%%%%%%%%%%%%%%%%%%%%%%%%%%%%%%%%%%%%%
\begin{acknowledgments}
The author is grateful to Dr. Amrita Mukherjee for the support regarding the computation.
\end{acknowledgments} 
%%%%%%%%%%%%%%%%%%%%%%%%%%%%%%%%%%%%%%%%%%%%%%%%%%%

%%%%%%%%%%%%%%%%%%%%%%%%%%%%%%%%%%%%%%%%%%%%%%%%%%%
\end{document}